\documentclass[twoside]{article}
\usepackage{amsmath, graphicx, amssymb, booktabs, multirow, makecell, qic}
\usepackage[caption=false]{subfig}

\newcommand{\braket}[2]{{\left\langle #1 \middle| #2 \right\rangle}}
\newcommand{\bra}[1]{{\left\langle #1 \right|}}
\newcommand{\ket}[1]{{\left| #1 \right\rangle}}
\newcommand{\ketbra}[2]{{\left| #1 \middle\rangle \middle \langle #2 \right|}}
\newcommand{\fref}[1]{Fig.~\ref{#1}}
\newcommand{\tref}[1]{Table~\ref{#1}}
\newcommand{\norm}[1]{\Vert #1 \Vert}
\newcommand{\ie}[1]{{\textit{i.e.}},}

\begin{document}

\setlength{\textheight}{8.0truein}    

\runninghead{Unstructured Search by Random and Quantum Walk}
            {T.~G.~Wong}

\normalsize\textlineskip
\thispagestyle{empty}
\setcounter{page}{1}

\copyrightheading{0}{0}{0000}{000--000}

\vspace*{0.88truein}

\alphfootnote

\fpage{1}

\centerline{\bf UNSTRUCTURED SEARCH BY RANDOM AND QUANTUM WALK}
\vspace*{0.37truein}
\centerline{\footnotesize THOMAS G.~WONG}
\vspace*{0.015truein}
\centerline{\footnotesize\it Department of Physics, Creighton University}
\baselineskip=10pt
\centerline{\footnotesize\it 2500 California Plaza, Omaha, Nebraska 68178, USA}
\vspace*{0.225truein}
\publisher{(received date)}{(revised date)}

\vspace*{0.21truein}

\abstracts{The task of finding an entry in an unsorted list of $N$ elements famously takes $O(N)$ queries to an oracle for a classical computer and $O(\sqrt{N})$ queries for a quantum computer using Grover's algorithm. Reformulated as a spatial search problem, this corresponds to searching the complete graph, or all-to-all network, for a marked vertex by querying an oracle. In this tutorial, we derive how discrete- and continuous-time (classical) random walks and quantum walks solve this problem in a thorough and pedagogical manner, providing an accessible introduction to how random and quantum walks can be used to search spatial regions. Some of the results are already known, but many are new. For large $N$, the random walks converge to the same evolution, both taking $N \ln(1/\epsilon)$ time to reach a success probability of $1-\epsilon$. In contrast, the discrete-time quantum walk asymptotically takes $\pi\sqrt{N}/2\sqrt{2}$ timesteps to reach a success probability of $1/2$, while the continuous-time quantum walk takes $\pi\sqrt{N}/2$ time to reach a success probability of $1$.}{}{}

\vspace*{10pt}

\vspace*{3pt}
\communicate{to be filled by the Editorial}

\vspace*{1pt}\textlineskip


\section{Introduction}

The motivation first given by Grover \cite{Grover1996} for his quantum algorithm for searching an unordered database was searching a telephone directory. Although the names are alphabetical, the phone numbers have no particular ordering. Then, to classically find a phone number in the list without its corresponding name, one can do no better than to go through each entry until the phone number is found. Checking whether a listed phone number matches the ``marked'' number we are searching for is akin to querying an oracle. Assume the marked entry is equally likely to be the first entry, the second entry, and so forth through the last entry. Then, if there are $N$ entries, the expected number of entries that must be examined is $(1 + 2 + \dots + N)/N = (N+1)/2 = O(N)$. Grover, however, introduced a quantum search algorithm that solves this problem in $O(\sqrt{N})$ queries by taking a uniform quantum superposition over the database, negating the amplitude of the marked entry by querying an oracle, inverting all the amplitudes about their average, and repeating the query and inversion until the probability of success was high. The runtime, or number of queries, is $O(\sqrt{N})$, and this scaling is optimal \cite{BBBV1997}.

Benioff \cite{Benioff2002} noted that Grover's algorithm, in order to invert each amplitude about the average, requires direct access to all entries of the database. If the database is physical, however, then it takes time to traverse the database, such as the time it takes to flip through the pages of a telephone book, since the entires are not all immediately accessible. This raises the question as to whether a quantum speedup is possible when searching spatial regions. This spatial search problem was investigated by Benioff in his original proposal of the problem \cite{Benioff2002}, by Aaronson and Ambainis through a recursive method \cite{AA2005}, and extensively via quantum walks \cite{CG2004,AKR2005,Chakraborty2016}.

\begin{figure}
\begin{center}
	\includegraphics{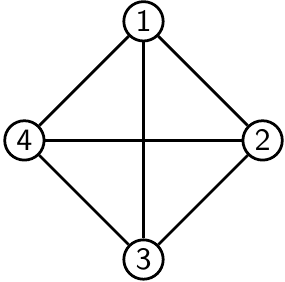}
	\vspace*{13pt}
	\fcaption{\label{fig:K4} The complete graph of $N = 4$ vertices.}
\end{center}
\end{figure}

The unstructured search problem, then, corresponds to spatial search on the complete graph, or all-to-all network, of $N$ vertices, since every element is accessible from every other. An example with $N = 4$ vertices is shown in \fref{fig:K4}. It is well-known in the quantum computing literature that the complete graph can be searched by a quantum walk in $O(\sqrt{N})$ time \cite{CG2004,Wong10}. More precisely, with a discrete-time quantum walk, the success probability reaches $1/2$ after $\pi\sqrt{N}/2\sqrt{2}$ timesteps \cite{Wong10}, and with a continuous-time quantum walk, the success probability reaches $1$ after $\pi\sqrt{N}/2$ time \cite{CG2004}. For (classical) random walks, it is known that their runtimes must be $O(N)$, since the quantum walk-based algorithms are quadratic speedups over the random walks \cite{Magniez2012,Krovi2016}. The precise details of the runtimes, however, are either absent from the literature or scantly known.

\begin{table}
\begin{center}
\tcaption{\label{table:summary}Summary of unstructured search by random and quantum walk.}
\rotatebox{90}{
\begin{tabular}{ccccc}
	\toprule
	\multirow{2}{*}{Property} & \multicolumn{2}{c}{Random Walk} & \multicolumn{2}{c}{Quantum Walk} \\
	& Discrete-Time & Continuous-Time & Discrete-Time & Continuous-Time \\
	\midrule
	Oracle & Absorbing & Absorbing & Phase & Hamiltonian Phase \\
	\midrule
	Vector Space & $\mathbb{R}^N$ & $\mathbb{R}^N$ & $\mathbb{C}^N \otimes \mathbb{C}^{N-1} \to \mathbb{R}^N \otimes \mathbb{R}^{N-1}$ & $\mathbb{C}^N$ \\
	\midrule
	Initial State & $\displaystyle \vec{p}(0) = \frac{1}{N} \sum_i \hat{e}_i$ & $\displaystyle \vec{p}(0) = \frac{1}{N} \sum_i \hat{e}_i$ & $\displaystyle \frac{1}{\sqrt{N(N-1)}} \sum_i \sum_{j \ne i} \ket{i \to j}$ & $\displaystyle \frac{1}{\sqrt{N}} \sum_i \ket{i}$ \\
	\midrule
	Evolution & $\vec{p}(t+1) = P \vec{p}(t)$ & $\displaystyle \frac{dp}{dt} = \frac{L}{\norm{L}} p$ & $\ket{\psi(t+1)} = SCQ \ket{\psi(t)}$ & $\displaystyle i \frac{d\ket{\psi}}{dt} = (-\gamma L - \ketbra{a}{a}) \ket{\psi}$ \\
	\midrule
	Solution & $\vec{p}(t) = P^t \vec{p}(0)$ & $\vec{p}(t) = e^{Lt/\norm{L}} \vec{p}(0)$ & $\ket{\psi(t)} = (SCQ)^t \ket{\psi(0)}$ & $\ket{\psi(t)} = e^{i(\gamma L + \ketbra{a}{a})t} \ket{\psi(0)}$ \\
	\midrule
	Subspace & $\mathbb{R}^2$ & $\mathbb{R}^2$ & $\mathbb{C}^3 \to \mathbb{R}^3$ & $\mathbb{C}^2$ \\
	\midrule
	\multirow{3}[4]{*}{\makecell{Success Probability \\ at Time $t$}} & \multirow{3}[4]{*}{$\displaystyle 1 - \frac{N-1}{N} \left( \frac{N-2}{N-1} \right)^t$} & \multirow{3}[4]{*}{$\displaystyle 1 - \frac{N-1}{N} e^{-t/N}$} & $\displaystyle \Big\{ (N-1) \left[ \cos(\phi t) + \sqrt{2N-3} \sin(\phi t) \right]$ & \multirow{3}[4]{*}{$\displaystyle \sin^2 \left( \frac{t}{\sqrt{N}} \right) + \frac{1}{N} \cos^2 \left( \frac{t}{\sqrt{N}} \right)$} \\
	& & & $\displaystyle + (-1)^t (N-2) \Big\}^2 / \left[ (2N-3)^2N \right]$, & \\
	& & & where $\displaystyle \phi = \sin^{-1} \left( \frac{\sqrt{2N-3}}{N-1} \right)$ & \\
	\midrule
	\multirow{2}[3]{*}{\makecell{Probability \\ and Runtime}} & $1 - \epsilon$ & $1 - \epsilon$ & $\displaystyle \frac{N(\sqrt{2N}+1)^2}{(2N-3)^2} + O \left( \frac{1}{N} \right)$ & 1 \\
	& $\log_{\frac{N-2}{N-1}} \left( \frac{N}{N-1} \epsilon \right)$ & $N \ln \left( \frac{N-1}{N} \frac{1}{\epsilon} \right)$ & $\displaystyle \frac{\pi}{2\phi}$ & $\displaystyle \frac{\pi}{2} \sqrt{N}$ \\
	\midrule
	\multirow{2}[0]{*}{\makecell{Asymptotic Prob. \\ and Runtime}} & $1 - \epsilon$ & $1 - \epsilon$ & \multirow{2}[0]{*}{\begin{tabular}{cc} $1/2$ & $1-\epsilon$ \\ $\displaystyle \frac{\pi}{2\sqrt{2}} \sqrt{N}$ & $\displaystyle \frac{\pi}{2\sqrt{2}} \sqrt{N} \log_2(1/\epsilon)$ \end{tabular}} & 1 \\
	& $N \ln(1/\epsilon)$ & $N \ln(1/\epsilon)$ & & $\displaystyle \frac{\pi}{2} \sqrt{N}$ \\
	\midrule
	Overall Runtime & $O(N)$ & $O(N)$ & $O(\sqrt{N})$ & $O(\sqrt{N})$ \\
	\bottomrule
\end{tabular}
}
\end{center}
\end{table}

In this tutorial, we fill this vacancy by deriving how both discrete- and continuous-time random walks search the complete graph for a marked vertex. While their exact evolutions differ, we prove that for large $N$, they converge to the same behavior, and they both reach a success probability of $1-\epsilon$ after $N \ln(1/\epsilon)$ time. Besides these seemingly new results, we also derive the behaviors of the quantum walks in order to provide a comparison. For the discrete-time quantum walk, the asymptotic behavior is known from \cite{Wong10}, but here we also derive the exact behavior. For the continuous-time quantum walk, the evolution is known from \cite{Wong10,CG2004,FG1998a}. Altogether, this tutorial provides an accessible introduction to how random and quantum walks solve spatial search problems. All of the results are summarized in \tref{table:summary}, and as we progress through the tutorial, we will explain each entry of the table.


\section{Discrete-Time Random Walk}

\begin{figure}
\begin{center}
	\includegraphics{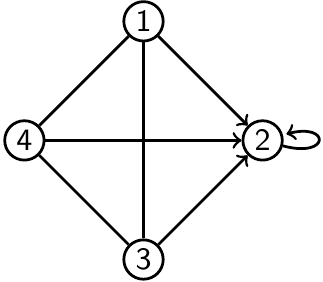}
	\vspace*{13pt}
	\fcaption{\label{fig:K4_absorbing} The complete graph of $N = 4$ vertices, where vertex $a = 2$ is classically marked by an absorbing vertex.}
\end{center}
\end{figure}

To solve the spatial search problem using a discrete-time random walk, we randomly jump from vertex to vertex, querying an oracle (or ``black box'') at each step to see if the current vertex is the one we are looking for. Let us label the marked vertex $a$. If we have found the marked vertex, we are done, so we stop. Then, the marked vertex is an absorbing vertex, and an example is shown in \fref{fig:K4_absorbing} with $N = 4$ vertices and $a = 2$ as the marked vertex. One can hop to the absorbing vertex, but once there, the walker stays put. That the oracle acts as an absorbing vertex is listed in the first row (not the titular row) of \tref{table:summary}.

To describe this process using linear algebra, we represent each vertex $1, 2, \dots, N$ by a basis column vector $\hat{e}_1, \hat{e}_2, \dots, \hat{e}_N$:
\[ \hat{e}_1 = \begin{pmatrix} 1 \\ 0 \\ 0 \\ \vdots \\ 0 \end{pmatrix}, \quad \hat{e}_2 = \begin{pmatrix} 0 \\ 1 \\ 0 \\ \vdots \\ 0 \end{pmatrix}, \quad \dots, \quad \hat{e}_N = \begin{pmatrix} 0 \\ 0 \\ 0 \\ \vdots \\ 1 \end{pmatrix}. \]
Then, the state of the random walk is a probability distribution over the vertices. That is, the probability distribution $\vec{p}$ can be written as
\[ \vec{p} = p_1 \hat{e}_1 + p_2 \hat{e}_2 + \dots + p_N \hat{e}_N = \begin{pmatrix} p_1 \\ p_2 \\ \vdots \\ p_N \end{pmatrix}, \]
where $p_i \in \mathbb{R}_{\ge 0}$, and since the total probability must be one, $p_1 + p_2 + \dots + p_N = 1$. So, the random walk occurs in the vector space $\mathbb{R}^N$. This is listed in the second row of \tref{table:summary}.

Each vertex has an equal probability of being the starting position of the walker, so the initial state $\vec{p}(0)$ is a uniform probability distribution over the vertices:
\begin{equation}
	\label{eq:p0}
	\vec{p}(0) = \frac{1}{N} \sum_{i = 1}^N \hat{e}_i = \frac{1}{N} \begin{pmatrix} 1 \\ 1 \\ \vdots \\ 1 \\ \end{pmatrix}. 
\end{equation}
This is listed in the third row of \tref{table:summary}. Note the total probability is given by summing all the entries of $\vec{p}(0)$, and it yields 1, as expected.

\begin{figure}
\begin{center}
	\includegraphics{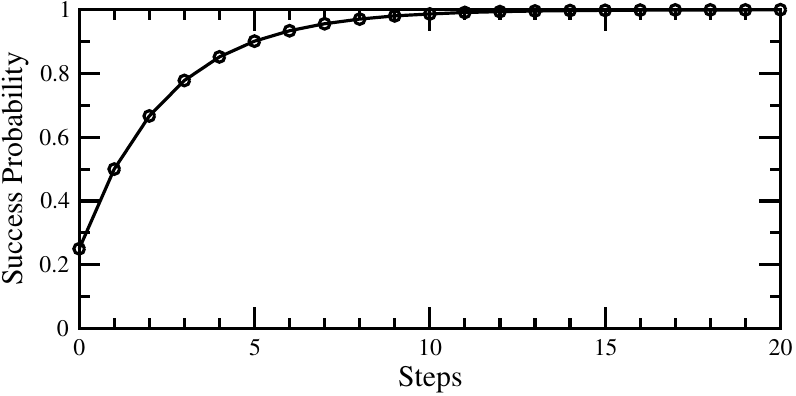}
	\vspace*{13pt}
	\fcaption{\label{fig:classical_discrete_N4} Success probability for search on the complete graph of $N = 4$ by discrete-time (classical) random walk.}
\end{center}
\end{figure}

A step of the random walk is obtained by multiplying the state on the left by a transition matrix
\[ P = \begin{pmatrix}
	P_{11} & P_{12} & \dots & P_{1N} \\
	P_{21} & P_{22} & \dots & P_{2N} \\
	\vdots & \vdots & \ddots & \vdots \\
	P_{N1} & P_{N2} & \dots & P_{NN} \\
\end{pmatrix}, \]
where $P_{ij}$ is the probability that the walker at vertex $j$ hops to vertex $i$, so the columns of $P$ must sum to 1 (\ie, the matrix is stochastic). For example, in \fref{fig:K4_absorbing} where $N = 4$ and $a = 2$, the transition matrix $P$ is
\[ \begin{pmatrix}
	          0 & 0 & \frac{1}{3} & \frac{1}{3} \\
	\frac{1}{3} & 1 & \frac{1}{3} & \frac{1}{3} \\
	\frac{1}{3} & 0 &           0 & \frac{1}{3} \\
	\frac{1}{3} & 0 & \frac{1}{3} &           0 \\
\end{pmatrix}. \]
As required, the sum of each column is 1. This transition matrix generalizes to $N$ vertices in a straightforward manner. For column $a$, which corresponds to the absorbing marked vertex, there is a $1$ on the diagonal and zeros elsewhere. Otherwise, the column corresponds to an unmarked vertex, and there is a zero on the diagonal and $1/(N-1)$ on the off-diagonals. That is, for our search algorithm,
\begin{equation}
	\label{eq:P}
	P_{ij} = \begin{cases}
		1, & \text{if }j = a, i = j, \\
		0, & \text{if }j = a, i \ne j, \\
		0, & \text{if }j \ne a, i = j, \\
		1/(N-1), & \text{if }j \ne a, i \ne j.
	\end{cases}
\end{equation}
Using the transition matrix, the state of the walker at time $t+1$ is obtained from the state at time $t$ via
\begin{equation}
	\label{eq:classical_discrete_evolution}
	\vec{p}(t+1) = P \vec{p}(t).
\end{equation}
This is listed in the fourth row of \tref{table:summary}. Since the initial state of the walker is $\vec{p}(0)$, the probability distribution after $t$ timesteps is
\begin{equation}
	\label{eq:classical_discrete_solution}
	\vec{p}(t) = P^t \vec{p}(0).
\end{equation}
This is listed in the fifth row of \tref{table:summary}. For example, when $N = 4$ and $a = 2$ as in \fref{fig:K4_absorbing}, we can use \eqref{eq:p0}, \eqref{eq:P}, and \eqref{eq:classical_discrete_evolution} or \eqref{eq:classical_discrete_solution} to numerically calculate the probability distribution at each timestep. From this, the success probability at each time (\ie, the probability at the marked/absorbing vertex) can be obtained, and we plot it in \fref{fig:classical_discrete_N4}. The initial success probability is $1/4 = 0.25$, and as steps of the random walk are taken, the probability builds up more and more at the marked vertex, asymptotically approaching $1$.

To analytically prove this behavior, note that due to the symmetry of the search problem, all the unmarked vertices evolve identically. Then, the system evolves in a 2D subspace spanned by $\{ \hat{e}_a, \hat{e}_b \}$, where $\hat{e}_b$ is the uniform distribution over the unmarked vertices:
\begin{equation}
	\label{eq:eb}
	\hat{e}_b = \frac{1}{N-1} \sum_{i \ne a} \hat{e}_i.
\end{equation}
Then, the initial uniform distribution \eqref{eq:p0} is
\[ \vec{p}(0) = \frac{1}{N} \sum_i \hat{e}_i = \frac{1}{N} \hat{e}_a + \frac{1}{N} \sum_{i \ne a} \hat{e}_i = \frac{1}{N} \hat{e}_a + \frac{N-1}{N} \hat{e}_b. \]
Or as a vector in the 2D basis $\hat{e}_a = \begin{pmatrix} 1 \\ 0 \end{pmatrix}$ and $\hat{e}_b = \begin{pmatrix} 0 \\ 1 \end{pmatrix}$,
\begin{equation}
	\label{eq:p0_2D}
	\vec{p}(0) = \frac{1}{N} \begin{pmatrix} 1 \\ N-1 \end{pmatrix}.
\end{equation}
Next, to determine the transition matrix $P$ in the $\{ \hat{e}_a, \hat{e}_b \}$ basis, we calculate how $P$ acts on $\hat{e}_a$ and $\hat{e}_b$. First,
\[ P \hat{e}_a = \hat{e}_a. \]
Next,
\begin{align*}
	P \hat{e}_b 
		&= \frac{1}{N-1} \sum_{i \ne a} P \hat{e}_i = \frac{1}{N-1} \sum_{i \ne a} \frac{1}{N-1} \sum_{j \ne i} e_j \\
		&= \frac{1}{(N-1)^2} \sum_{i \ne a} \left( \hat{e}_a + \sum_{\substack{j \ne i \\j \ne a}} \hat{e}_j \right) = \frac{1}{(N-1)^2} \left( \sum_{i \ne a} \hat{e}_a + \sum_{i \ne a} \sum_{\substack{j \ne i \\j \ne a}} \hat{e}_j \right) \\
		&= \frac{1}{(N-1)^2} \left[ (N-1) \hat{e}_a + (N-2) \sum_{i \ne a} \hat{e}_i \right] = \frac{1}{N-1} \hat{e}_a + \frac{N-2}{N-1} \hat{e}_b,
\end{align*}
where to get the third line, we used $\sum_{i \ne a} \sum_{j \ne i, j \ne a} \hat{e}_j = (N-2) \sum_{i \ne a} \hat{e}_i$ because each unmarked vertex appears once from each of the $N-2$ other unmarked vertices. So, as a matrix in the $\{ \hat{e}_a, \hat{e}_b \}$ basis,
\begin{equation}
	\label{eq:P_2D}
	P = \begin{pmatrix}
		1 & \frac{1}{N-1} \\
		0 & \frac{N-2}{N-1} \\
	\end{pmatrix}.
\end{equation}
The system still evolves according to \eqref{eq:classical_discrete_solution}, but now we can use the two-dimensional initial state \eqref{eq:p0_2D} and two-dimensional transition matrix \eqref{eq:P_2D}. That the system evolves in the subspace $\mathbb{R}^2$ is listed in the sixth row of \tref{table:summary}. To derive the behavior of the algorithm, we can find the eigenvectors and eigenvalues of \eqref{eq:P_2D}. They are
\begin{alignat*}{3}
	\vec{v}_1 &= \begin{pmatrix}
		-1 \\
		1 \\
	\end{pmatrix} = -\hat{e}_a + \hat{e}_b, \quad &&\lambda_1 = \frac{N-2}{N-1}, \\
	\vec{v}_2 &= \begin{pmatrix}
		1 \\
		0 \\
	\end{pmatrix} = \hat{e}_a, &&\lambda_2 = 1.
\end{alignat*}
Then, the initial state \eqref{eq:p0_2D} can be expressed as a linear combination of these eigenvectors:
\[ \vec{p}(0) = \frac{N-1}{N} \vec{v}_1 + \vec{v}_2. \]
Thus, using \eqref{eq:classical_discrete_solution}, the probability distribution of the random walk at time $t$ is
\begin{align*}
	\vec{p}(t)
		&= P^t \vec{p}(0) \\
		&= \frac{N-1}{N} P^t \vec{v}_1 + P^t \vec{v}_2 \\
		&= \frac{N-1}{N} \lambda_1^t \vec{v}_1 + \lambda_2^t \vec{v}_2 \\
		&= \frac{N-1}{N} \left( \frac{N-2}{N-1} \right)^t \left( -\hat{e}_a + \hat{e}_b \right) + 1^t \hat{e}_a \\
		&= \left[ 1 - \frac{N-1}{N} \left( \frac{N-2}{N-1} \right)^t \right] \hat{e}_a + \frac{N-1}{N} \left( \frac{N-2}{N-1} \right)^t \hat{e}_b. 
\end{align*}
The success probability at time $t$ is simply the coefficient of $\hat{e}_a$:
\begin{equation}
	\label{eq:classical_discrete_prob_time}
	1 - \frac{N-1}{N} \left( \frac{N-2}{N-1} \right)^t.
\end{equation}
This formula exactly reproduces \fref{fig:classical_discrete_N4} with $N = 4$, and as far as we know, it is a new result. It is listed in the seventh row of \tref{table:summary}.

There is another way to arrive at this result using basic probability theory. In this approach, we first find the probability of \emph{not} finding the marked vertex at time $t$. Then, the probability of finding the marked vertex at time $t$ is 1 minus this. To begin, at $t = 0$, we start at a random vertex, so the probability of \emph{not} finding the marked vertex is $(N-1)/N$. Then, taking a step of the random walk, the probability of continuing to not find the marked vertex is obtained by multiplying $(N-1)/N$ by $(N-2)/(N-1)$ because the walker can jump to $N-1$ other vertices, $N-2$ of which are unmarked. With each step of the quantum walk, we continue to multiply by $(N-2)/(N-1)$. Thus, after $t$ steps of the random walk, the probability that the marked vertex has not been found is
\[ \frac{N-1}{N} \left( \frac{N-2}{N-1} \right)^t. \]
Finally, taking the complement, the probability of finding the marked vertex after $t$ steps is
\[ 1 - \frac{N-1}{N} \left( \frac{N-2}{N-1} \right)^t, \]
which is the same as \eqref{eq:classical_discrete_prob_time}. While this approach is shorter, it may not work for other graphs. In contrast, our first approach of carefully analyzing the random walk using linear algebra is more general, and the quantum walk can be analyzed using linear algebra in a similar manner.

Since the success probability asymptotically approaches 1, it would take infinite time for the success probability to reach 1. So, for the runtime of the algorithm, we instead calculate the time it takes the success probability reach a value of $1-\epsilon$. To find this, we simply set \eqref{eq:classical_discrete_prob_time} equal to $1-\epsilon$ and solve for $t$:
\begin{gather}
	1 - \frac{N-1}{N} \left( \frac{N-2}{N-1} \right)^t = 1 - \epsilon \nonumber \\
	\frac{N-1}{N} \left( \frac{N-2}{N-1} \right)^t = \epsilon \nonumber \\
	\left( \frac{N-2}{N-1} \right)^t = \frac{N}{N-1} \epsilon \nonumber \\
	t = \log_{\frac{N-2}{N-1}} \left( \frac{N}{N-1} \epsilon \right) \label{eq:classical_discrete_runtime}.
\end{gather}
This is the runtime, and it seems to be a new result. It, along with its corresponding success probability of $1-\epsilon$, are listed in the eighth row of \tref{table:summary}.

For large $N$, we can simplify this runtime. First, we rewrite it using the fact that $\log_b a = (\log_c a) / (\log_c b)$ for any base $c$. Choosing $c = e$ so that we use the natural logarithm $\ln = \log_e$, the runtime is
\[ t = \frac{\ln \left( \frac{N}{N-1} \epsilon \right)}{\ln \left( \frac{N-2}{N-1} \right)}. \]
Now, we can Taylor expand the numerator and denominator for small $\epsilon$ and large $N$. The numerator is
\[ \ln \left( \frac{N}{N-1} \epsilon \right) \approx \ln \epsilon, \]
and the denominator is
\[ \ln \left( \frac{N-2}{N-1} \right) = \ln \left( 1 - \frac{1}{N-1} \right) \approx \ln \left( 1 - \frac{1}{N} \right) \approx \frac{-1}{N}. \]
Thus, the runtime is asymptotically
\begin{equation}
	\label{eq:classical_discrete_runtime_largeN}
	t \approx -N \ln \epsilon = N \ln \left( \frac{1}{\epsilon} \right).
\end{equation}
This asymptotic runtime seems to be a new result, and it is listed in the ninth row of \tref{table:summary}.

If $\epsilon$ is a constant, then the runtime $N \ln(1/\epsilon) = O(N)$, and the success probability $1-\epsilon = O(1)$. Even though the success probability is less than 1, since it is a constant, we expect to repeat the algorithm only a constant number of times, on average, so the overall runtime is $O(N)$, and this is listed in the final row of \tref{table:summary}. This overall runtime of $O(N)$ is expected because it is known that the random walk is quadratically slower than the quantum walk's $O(\sqrt{N})$ time, which we will review later.


\section{Continuous-Time Random Walk}

A continuous-time random walk bears many similarities to the discrete-time random walk we just investigated. The walker still hops from vertex to vertex on the complete graph, querying an oracle as it goes, and stopping when it has found the marked vertex. So, the marked vertex still acts as an absorbing vertex, as listed in the first row of \tref{table:summary}. The state of the system is still a probability distribution over the $N$ vertices, so it evolves in the vector space $\mathbb{R}^N$, as listed in the second row of \tref{table:summary}. Again, each vertex is represented by an orthonormal basis vector $\hat{e}_1, \hat{e}_2, \dots, \hat{e}_N$. As before, the initial state $\vec{p}(0)$ is a uniform distribution over the vertices \eqref{eq:p0}, as listed in the third row of \tref{table:summary}, but now it evolves in continuous-time by a differential equation:
\begin{equation}
	\label{eq:classical_continuous_evolution}
	\frac{d\vec{p}}{dt} = \frac{L}{\norm{L}} \vec{p},
\end{equation}
where $L = A - D$ is the spatially discrete Laplacian defined by the adjacency matrix $A$, where $A_{ij} = 1$ if there is an edge from vertex $j$ to vertex $i$ and $A_{ij} = 0$ otherwise, and the diagonal degree matrix $D$, where $D_{ii} = \text{deg}(i)$ is the out-degree of vertex $i$. $\norm{L}$ denotes the spectral norm of $L$, and it appears in \eqref{eq:classical_continuous_evolution} so that the probability per unit time of a transition occurring remains constant in the number of vertices $N$. This differential equation is listed in the fourth row of \tref{table:summary}, and its solution is
\begin{equation}
	\label{eq:classical_continuous_solution}
	\vec{p}(t) = e^{Lt / \norm{L}} \vec{p}(0).
\end{equation}
This solution is listed in the fifth row of \tref{table:summary}. Note the columns of $L$ sum to zero, and this ensures that the total probability remains $1$.

Note in spectral graph theory, it is common for the combinatorial Laplacian $\mathcal{L} = D - A$ to be used instead of the discrete Laplacian $L = A - D$, since $\mathcal{L}$ is positive-semidefinite, meaning the eigenvalues of $\mathcal{L}$ are non-negative. In this tutorial, we use $L = A - D$ because it is a discretization of the Laplace operator, which in Cartesian coordinates is $\nabla^2 = \partial^2/\partial x^2 + \partial^2/\partial y^2 + \partial^2/\partial z^2$. Then, \eqref{eq:classical_continuous_evolution} takes the form of a discrete heat or diffusion equation, as expected for a randomly walking particle.

\begin{figure}
\begin{center}
	\includegraphics{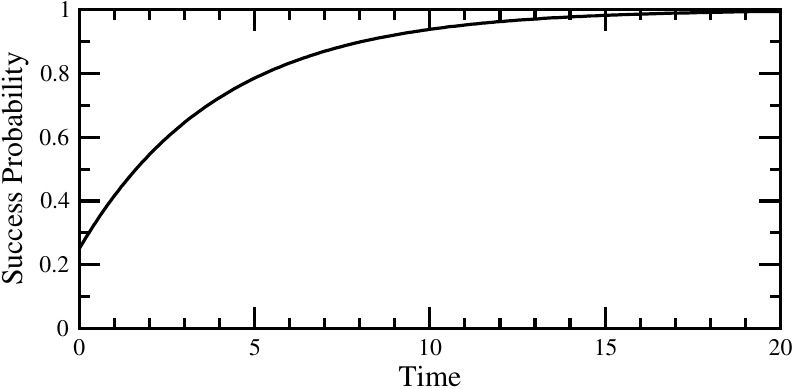}
	\vspace*{13pt}
	\fcaption{\label{fig:classical_continuous_N4} Success probability for search on the complete graph of $N = 4$ by continuous-time (classical) random walk.}
\end{center}
\end{figure}

For example, in \fref{fig:K4_absorbing}, where $N = 4$ and $a = 2$, we have
\[ A = \begin{pmatrix}
	0 & 0 & 1 & 1 \\
	1 & 1 & 1 & 1 \\
	1 & 0 & 0 & 1 \\
	1 & 0 & 1 & 0 \\
\end{pmatrix}, \enspace
D = \begin{pmatrix}
	3 & 0 & 0 & 0 \\
	0 & 1 & 0 & 0 \\
	0 & 0 & 3 & 0 \\
	0 & 0 & 0 & 3 \\
\end{pmatrix}, \enspace
L = \begin{pmatrix}
	-3 & 0 &  1 &  1 \\
	 1 & 0 &  1 &  1 \\
	 1 & 0 & -3 &  1 \\
	 1 & 0 &  1 & -3 \\
\end{pmatrix}, \enspace \norm{L} = 4. \]
Using these, we can numerically calculate the probability distribution of the walker using \eqref{eq:classical_continuous_solution}, and we plot the success probability vs time in \fref{fig:classical_continuous_N4}. It starts at $1/4 = 0.25$, and as expected, as the walk progresses, the probability builds up at the marked absorbing vertex.

For general $N$ and $a$, the elements of $L$ are
\begin{equation}
	\label{eq:classical_L}
	L_{ij} = \begin{cases}
		0, & \text{if } j = a, \\
		-(N-1), & \text{if } j \ne a, i = j, \\
		1, & \text{if } j \ne a, i \ne j. \\
	\end{cases}
\end{equation}
$L$ is a symmetric matrix ($L_{ij} = L_{ji}$), and the spectral norm of a symmetric matrix equals its largest eigenvalue in absolute value. Then, since $L$ has eigenvalues 0 and $-N$ (with multiplicity $N-1$), its spectral norm is
\[ \norm{L} = N. \]

We can analytically prove the behavior of this algorithm by again working in the 2D subspace with orthonormal basis vectors $\hat{e}_a$ and $\hat{e}_b$ \eqref{eq:eb}. This is listed in the sixth row of \tref{table:summary}. The initial uniform distribution $\vec{p}(0)$ is again given by \eqref{eq:p0_2D}. To determine the discrete Laplacian in the $\{ \hat{e}_a, \hat{e}_b \}$ basis, we calculate how it acts on each basis vector:
\[ L \hat{e}_a = 0, \]
and
\begin{align*}
	L \hat{e}_b 
		&= \frac{1}{N-1} \sum_{i \ne a} L \hat{e}_i = \frac{1}{N-1} \sum_{i \ne a} \left( \hat{e}_a - (N-1) \hat{e}_i + \sum_{\substack{j \ne i \\ j \ne a}} \hat{e}_j \right) \\
		&= \hat{e}_a - (N-1) \hat{e}_b + \frac{1}{N-1} \sum_{i \ne a} \sum_{\substack{j \ne i \\ j \ne a}} \hat{e}_j \\
		&= \hat{e}_a - (N-1) \hat{e}_b + \frac{N-2}{N-1} \sum_{i \ne a} \hat{e}_i \\
		&= \hat{e}_a - (N-1) \hat{e}_b + (N-2) \hat{e}_b \\
		&= \hat{e}_a - \hat{e}_b.
\end{align*}
So, in the $\{ \hat{e}_a, \hat{e}_b \}$ subspace,
\[ L = \begin{pmatrix}
	0 & 1 \\
	0 & -1 \\
\end{pmatrix}. \]
The eigenvectors and eigenvalues of this are
\begin{alignat*}{3}
	\vec{v}_1 &= \begin{pmatrix}
		-1 \\
		1 \\
	\end{pmatrix} = -\hat{e}_a + \hat{e}_b, \quad &&\lambda_1 = -1, \\
	\vec{v}_2 &= \begin{pmatrix}
		1 \\
		0 \\
	\end{pmatrix} = \hat{e}_a, \quad &&\lambda_2 = 0.
\end{alignat*}
Then, the initial state can be expressed as a linear combination of these eigenvectors:
\[ \vec{p}(0) = \frac{N-1}{N} \vec{v}_1 + \vec{v}_2. \]
Thus, using \eqref{eq:classical_continuous_solution}, the probability distribution at time $t$ is
\begin{align*}
	\vec{p}(t)
		&= e^{L t/\norm{L}} \vec{p}(0) \\
		&= \frac{N-1}{N} e^{\lambda_1 t/N} \vec{v}_1 + e^{\lambda_2 t/N} \vec{v}_2 \\
		&= \frac{N-1}{N} e^{-t/N} \left( -\hat{e}_a + \hat{e}_b \right) + e^{0} \hat{e}_a \\
		&= \left( 1 - \frac{N-1}{N} e^{-t/N} \right) \hat{e}_a + \frac{N-1}{N} e^{-t/N} \hat{e}_b.
\end{align*}
The success probability at time $t$ is simply the coefficient of $\hat{e}_a$:
\begin{equation}
	\label{eq:classical_continuous_prob_time}
	1 - \frac{N-1}{N} e^{-t/N}.
\end{equation}
This exactly reproduces \fref{fig:classical_continuous_N4} when $N = 4$, and it appears to be a new result. It is listed in the seventh row of \tref{table:summary}.

To determine the time at which the success probability reaches $1-\epsilon$, we set \eqref{eq:classical_continuous_prob_time} equal to $1-\epsilon$ and solve for $t$:
\begin{gather}
	1 - \frac{N-1}{N} e^{-t/N} = 1 - \epsilon \nonumber \\
	\frac{N-1}{N} e^{-t/N} = \epsilon \nonumber \\
	e^{-t/N} = \frac{N}{N-1} \epsilon \nonumber \\
	\frac{-t}{N} = \ln \left( \frac{N}{N-1} \epsilon \right) \nonumber \\
	\frac{t}{N} = \ln \left( \frac{N-1}{N} \frac{1}{\epsilon} \right) \nonumber \\
	t = N \ln \left( \frac{N-1}{N} \frac{1}{\epsilon} \right). \label{eq:classical_continuous_runtime}
\end{gather}
This is the exact runtime of the algorithm, and it seems to be a new result. It is listed in the eighth row of \tref{table:summary}, along with its corresponding success probability of $1-\epsilon$.

Now, for large $N$, this runtime becomes
\begin{equation}
	\label{eq:classical_continuous_runtime_largeN}
	t \approx N \ln \left( \frac{1}{\epsilon} \right).
\end{equation}
This is the same asymptotic, linear runtime as the discrete-time random walk, and it seems to be a new result. It is listed in the ninth row of \tref{table:summary}, and overall, the runtime is $O(N)$, as listed in the final row of \tref{table:summary}.

\begin{figure}
\begin{center}
	\subfloat[] {
		\includegraphics{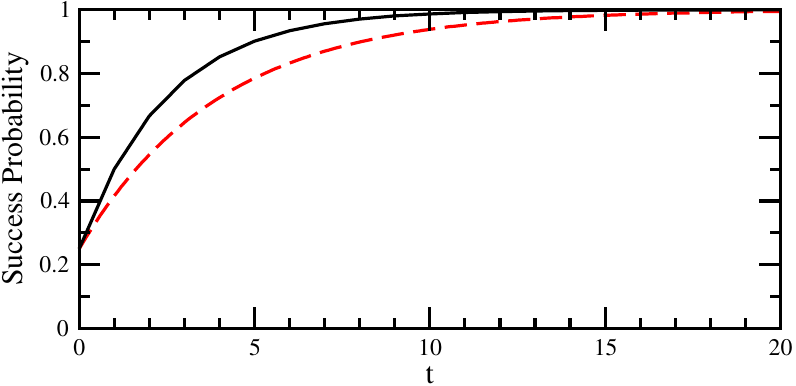}
		\label{fig:classical_N4}
	}

	\subfloat[] {
		\includegraphics{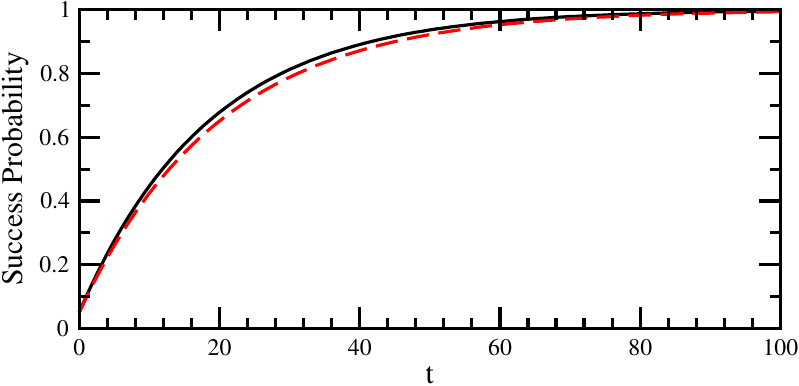}
		\label{fig:classical_N20}
	}

	\subfloat[] {
		\includegraphics{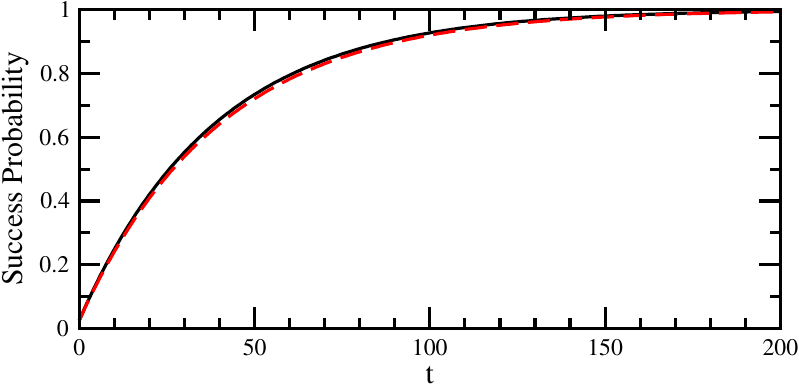}
		\label{fig:classical_N40}
	}
	\vspace*{13pt}
	\fcaption{\label{fig:classical_N4_N20_N40} Success probability for search on the complete graph of (a) $N = 4$, (b) $N = 20$, and $N = 40$ vertices, using a discrete-time random walk in solid black and a continuous-time random walk in dashed red.}
\end{center}
\end{figure}

\begin{figure}
\begin{center}
	\includegraphics{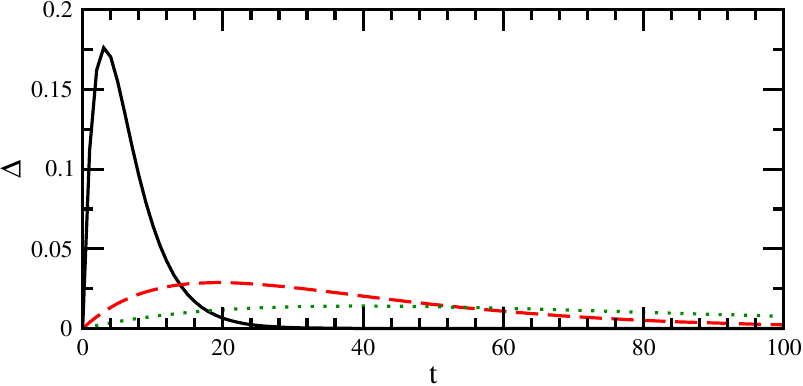}
	\vspace*{13pt}
	\fcaption{\label{fig:classical_Delta} $\mathrm{\Delta}$ as a function of $t$, where the solid black curve is $N = 4$, the dashed red curve is $N = 20$, and the dotted green curve is $N = 40$.}
\end{center}
\end{figure}

Let us also take a moment to compare the evolutions of the discrete-time random walk in \eqref{eq:classical_discrete_prob_time} and the continuous-time random walk in \eqref{eq:classical_continuous_prob_time}, \ie, the seventh row of \tref{table:summary}. For search on the complete graph of $N = 4$ vertices, both probabilities are plotted in \fref{fig:classical_N4}, with the discrete-time walk in solid black and the continuous-time walk in dashed-red. That is, this plot just combines \fref{fig:classical_discrete_N4} and \fref{fig:classical_continuous_N4}. Both success probabilities monotonically increase toward 1, but their curves are different. As $N$ increases, however, their behaviors converge, as shown in \fref{fig:classical_N20} with $N = 20$ and \fref{fig:classical_N40} with $N = 40$. We will prove this analytically now. Comparing \eqref{eq:classical_discrete_prob_time} and \eqref{eq:classical_continuous_prob_time}, the evolutions differ by a single term: The former has $[(N-2)/(N-1)]^t$, while the latter has $e^{-t/N}$. The difference between these is
\[ \mathrm{\Delta} = e^{-t/N} - \left( \frac{N-2}{N-1} \right)^t. \]
This difference is plotted in \fref{fig:classical_Delta} for $N = 4$, $20$, and $40$, and we see that the difference gets smaller as $N$ increases, meaning the evolutions of the discrete- and continuous-time random walks converge for large $N$. We can prove this by finding the maximum value of $\mathrm{\Delta}$ and showing that it goes to zero as $N \to \infty$. To find the maximum value of $\mathrm{\Delta}$, we use elementary calculus. The derivative of $\mathrm{\Delta}$ with respect to $t$ is
\[ \frac{d\mathrm{\Delta}}{dt} = \frac{-1}{N} e^{-t/N} - \left( \frac{N-2}{N-1} \right)^t \ln \left( \frac{N-2}{N-1} \right). \]
Setting this equal to $0$ and solving for $t$, $\mathrm{\Delta}$ is maximized when $t$ equals
\[ \frac{N \ln \left[ N \ln \left( \frac{N-1}{N-2} \right) \right]}{N \ln \left( \frac{N-1}{N-2} \right) - 1 }. \]
Plugging this value of $t$ into $\mathrm{\Delta}$, the maximum value of $\mathrm{\Delta}$ is
\[ \mathrm{\Delta}_\text{max} = e^{\frac{-\ln \left[ N \ln \left( \frac{N-1}{N-2} \right) \right]}{N \ln \left( \frac{N-1}{N-2} \right) - 1 }} - \left( \frac{N-2}{N-1} \right)^{\frac{N \ln \left[ N \ln \left( \frac{N-1}{N-2} \right) \right]}{N \ln \left( \frac{N-1}{N-2} \right) - 1 }}. \]
Taylor expanding this for large $N$,
\[ \mathrm{\Delta}_\text{max} = \frac{3}{2e N} + O\left( \frac{1}{N^2} \right). \]
This goes to zero as $N$ gets larger, so the difference between \eqref{eq:classical_discrete_prob_time} and \eqref{eq:classical_continuous_prob_time} goes to zero as $N$ gets larger, and thus the behaviors of the discrete- and continuous-time random walks converge for large $N$.


\section{Discrete-Time Quantum Walk}

In the previous two sections, we investigated how discrete- and continuous-time (classical) random walks search the complete graph for a marked vertex. In this section and the next, we explore how quantum walks solve this problem, beginning with the discrete-time quantum walk in this section and then the continuous-time quantum walk in the next section.

\begin{figure}
\begin{center}
	\subfloat[] {
		\includegraphics{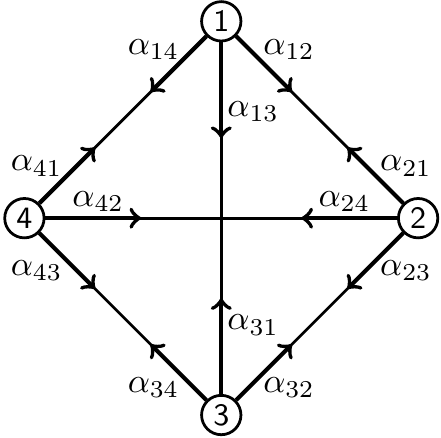}
		\label{fig:K4_directions} 
	} \quad \quad \quad
	\subfloat[] {
		\includegraphics{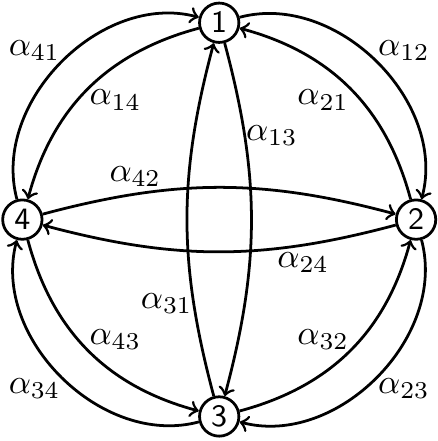}
		\label{fig:K4_arcs} 
	}
	\vspace*{13pt}
	\fcaption{The amplitudes of a discrete-time quantum walk on the complete graph of $N = 4$ vertices as (a) vertices with directions or (b) directed edges or arcs.}
\end{center}
\end{figure}

The discrete-time quantum walk \cite{Aharonov2001} differs from the discrete-time random walk in some key ways, which we introduce here before formalizing more precisely. First, besides the position of the walker, Meyer \cite{Meyer1996a,Meyer1996b} showed that for the quantum walk to evolve nontrivially, it also needs an internal degree of freedom that can be interpreted as the direction that the walker points. For example, for a (classical) random walk on the complete graph of $N = 4$ vertices (\fref{fig:K4}), there are only four numbers $p_1$, $p_2$, $p_3$, and $p_4$ corresponding to the probability at each vertex, and the state of the walk would be
\[ \vec{p} = \begin{pmatrix} p_1 \\ p_2 \\ p_3 \\ p_4 \end{pmatrix}. \]
In contrast, the discrete-time quantum walk on the complete graph of $N = 4$ vertices is depicted in \fref{fig:K4_directions}, and it has twelve numbers $\alpha_{12}, \dots, \alpha_{43}$, where $\alpha_{ij}$ corresponds to a walker at vertex $i$ pointing to vertex $j$. Then, the state of the quantum walk is
\begin{equation}
	\label{eq:quantum_discrete_psi_N4}
	\ket{\psi} = \begin{pmatrix} \alpha_{12} \\ \alpha_{13} \\ \alpha_{14} \\ \alpha_{21} \\ \alpha_{23} \\ \alpha_{24} \\ \alpha_{31} \\ \alpha_{32} \\ \alpha_{34} \\ \alpha_{41} \\ \alpha_{42} \\ \alpha_{43} \end{pmatrix}.
\end{equation}
Note in quantum mechanics, a generic quantum state is typically named $\psi$, and it is customary to use ``Dirac notation'' (also known as ``bra-ket notation''), where a column vector is written as a ``ket,'' which has a vertical bar on the left and a right angle on the right, \ie, $\ket{\psi}$. Its conjugate transpose, which is a row vector, is written as a ``bra,'' which has a left angle on the left and a vertical bar on the right, \ie, $\bra{\psi}$. Second, in quantum mechanics, instead of a state being a probability distribution, or nonnegative real-valued linear combination of the basis states, it is a ``superposition,'' or complex linear combination of the basis states. That is, the numbers $\alpha_{ij}$ are generally complex, and we call them ``amplitudes.'' Then, from the ``Born rule'' of quantum mechanics, when the walker is measured or observed, the probability that the walker is found at vertex $i$ pointing to vertex $j$ is equal to $|\alpha_{ij}|^2$. Thus, the probability that the walker is found at vertex $i$ is the sum of the norm-square of the amplitudes at $i$, or $\sum_{j \ne i} |\alpha_{ij}|^2$. For example, in \fref{fig:K4_directions}, the probability of finding the walker at vertex $1$ is $|\alpha_{12}|^2 + |\alpha_{13}|^2 + |\alpha_{14}|^2$. Since the total probability must be 1, $\sum_i \sum_{j \ne i} |\alpha_{ij}|^2 = 1$. Finally, as shown in \fref{fig:K4_arcs}, the quantum walk can be equivalently interpreted as occurring on the directed edges or arcs of the graph, since they take into account the directions that the walker can point.

\begin{table}
\begin{center}
\tcaption{\label{table:quantum_discrete}The amplitudes of a discrete-time quantum walk searching the complete graph of $N = 4$ vertices with $a = 2$ marked, for three timesteps.}
\rotatebox{90}{%
\begin{tabular}{r|ccc|ccc|ccc|ccc}
	\toprule
	& $\alpha_{12}$ & $\alpha_{13}$ & $\alpha_{14}$ & $\alpha_{21}$ & $\alpha_{23}$ & $\alpha_{24}$ & $\alpha_{31}$ & $\alpha_{32}$ & $\alpha_{34}$ & $\alpha_{41}$ & $\alpha_{42}$ & $\alpha_{43}$ \\
	\midrule
	$\ket{\psi(0)}$ & $\frac{1}{2\sqrt{3}}$ & $\frac{1}{2\sqrt{3}}$ & $\frac{1}{2\sqrt{3}}$ & $\frac{1}{2\sqrt{3}}$ & $\frac{1}{2\sqrt{3}}$ & $\frac{1}{2\sqrt{3}}$ & $\frac{1}{2\sqrt{3}}$ & $\frac{1}{2\sqrt{3}}$ & $\frac{1}{2\sqrt{3}}$ & $\frac{1}{2\sqrt{3}}$ & $\frac{1}{2\sqrt{3}}$ & $\frac{1}{2\sqrt{3}}$ \\[0.7em]
	Vertex Probability & & $\frac{1}{4}$ & & & $\frac{1}{4}$ & & & $\frac{1}{4}$ & & & $\frac{1}{4}$ & \\[0.25em]
	\midrule
	$Q\ket{\psi(0)}$ & $\frac{1}{2\sqrt{3}}$ & $\frac{1}{2\sqrt{3}}$ & $\frac{1}{2\sqrt{3}}$ & $\frac{-1}{2\sqrt{3}}$ & $\frac{-1}{2\sqrt{3}}$ & $\frac{-1}{2\sqrt{3}}$ & $\frac{1}{2\sqrt{3}}$ & $\frac{1}{2\sqrt{3}}$ & $\frac{1}{2\sqrt{3}}$ & $\frac{1}{2\sqrt{3}}$ & $\frac{1}{2\sqrt{3}}$ & $\frac{1}{2\sqrt{3}}$ \\[0.7em]
	$CQ\ket{\psi(0)}$ & $\frac{1}{2\sqrt{3}}$ & $\frac{1}{2\sqrt{3}}$ & $\frac{1}{2\sqrt{3}}$ & $\frac{-1}{2\sqrt{3}}$ & $\frac{-1}{2\sqrt{3}}$ & $\frac{-1}{2\sqrt{3}}$ & $\frac{1}{2\sqrt{3}}$ & $\frac{1}{2\sqrt{3}}$ & $\frac{1}{2\sqrt{3}}$ & $\frac{1}{2\sqrt{3}}$ & $\frac{1}{2\sqrt{3}}$ & $\frac{1}{2\sqrt{3}}$ \\[0.7em]
	$\ket{\psi(1)} = SCQ\ket{\psi(0)}$ & $\frac{-1}{2\sqrt{3}}$ & $\frac{1}{2\sqrt{3}}$ & $\frac{1}{2\sqrt{3}}$ & $\frac{1}{2\sqrt{3}}$ & $\frac{1}{2\sqrt{3}}$ & $\frac{1}{2\sqrt{3}}$ & $\frac{1}{2\sqrt{3}}$ & $\frac{-1}{2\sqrt{3}}$ & $\frac{1}{2\sqrt{3}}$ & $\frac{1}{2\sqrt{3}}$ & $\frac{-1}{2\sqrt{3}}$ & $\frac{1}{2\sqrt{3}}$ \\[0.7em]
	Vertex Probability & & $\frac{1}{4}$ & & & $\frac{1}{4}$ & & & $\frac{1}{4}$ & & & $\frac{1}{4}$ & \\[0.25em]
	\midrule
	$Q\ket{\psi(1)}$ & $\frac{-1}{2\sqrt{3}}$ & $\frac{1}{2\sqrt{3}}$ & $\frac{1}{2\sqrt{3}}$ & $\frac{-1}{2\sqrt{3}}$ & $\frac{-1}{2\sqrt{3}}$ & $\frac{-1}{2\sqrt{3}}$ & $\frac{1}{2\sqrt{3}}$ & $\frac{-1}{2\sqrt{3}}$ & $\frac{1}{2\sqrt{3}}$ & $\frac{1}{2\sqrt{3}}$ & $\frac{-1}{2\sqrt{3}}$ & $\frac{1}{2\sqrt{3}}$ \\[0.7em]
	$CQ\ket{\psi(1)}$ & $\frac{5}{6\sqrt{3}}$ & $\frac{-1}{6\sqrt{3}}$ & $\frac{-1}{6\sqrt{3}}$ & $\frac{-1}{2\sqrt{3}}$ & $\frac{-1}{2\sqrt{3}}$ & $\frac{-1}{2\sqrt{3}}$ & $\frac{-1}{6\sqrt{3}}$ & $\frac{5}{6\sqrt{3}}$ & $\frac{-1}{6\sqrt{3}}$ & $\frac{-1}{6\sqrt{3}}$ & $\frac{5}{6\sqrt{3}}$ & $\frac{-1}{6\sqrt{3}}$ \\[0.7em]
	$\ket{\psi(2)} = SCQ\ket{\psi(1)}$ & $\frac{-1}{2\sqrt{3}}$ & $\frac{-1}{6\sqrt{3}}$ & $\frac{-1}{6\sqrt{3}}$ & $\frac{5}{6\sqrt{3}}$ & $\frac{5}{6\sqrt{3}}$ & $\frac{5}{6\sqrt{3}}$ & $\frac{-1}{6\sqrt{3}}$ & $\frac{-1}{2\sqrt{3}}$ & $\frac{-1}{6\sqrt{3}}$ & $\frac{-1}{6\sqrt{3}}$ & $\frac{-1}{2\sqrt{3}}$ & $\frac{-1}{6\sqrt{3}}$ \\[0.7em]
	Vertex Probability & & $\frac{11}{108}$ & & & $\frac{25}{36}$ & & & $\frac{11}{108}$ & & & $\frac{11}{108}$ & \\[0.25em]
	\midrule
	$Q\ket{\psi(2)}$ & $\frac{-1}{2\sqrt{3}}$ & $\frac{-1}{6\sqrt{3}}$ & $\frac{-1}{6\sqrt{3}}$ & $\frac{-5}{6\sqrt{3}}$ & $\frac{-5}{6\sqrt{3}}$ & $\frac{-5}{6\sqrt{3}}$ & $\frac{-1}{6\sqrt{3}}$ & $\frac{-1}{2\sqrt{3}}$ & $\frac{-1}{6\sqrt{3}}$ & $\frac{-1}{6\sqrt{3}}$ & $\frac{-1}{2\sqrt{3}}$ & $\frac{-1}{6\sqrt{3}}$ \\[0.7em]
	$CQ\ket{\psi(2)}$ & $\frac{-1}{18\sqrt{3}}$ & $\frac{-7}{18\sqrt{3}}$ & $\frac{-7}{18\sqrt{3}}$ & $\frac{-5}{6\sqrt{3}}$ & $\frac{-5}{6\sqrt{3}}$ & $\frac{-5}{6\sqrt{3}}$ & $\frac{-7}{18\sqrt{3}}$ & $\frac{-1}{18\sqrt{3}}$ & $\frac{-7}{18\sqrt{3}}$ & $\frac{-7}{18\sqrt{3}}$ & $\frac{-1}{18\sqrt{3}}$ & $\frac{-7}{6\sqrt{3}}$ \\[0.7em]
	$\ket{\psi{(3)}} = SCQ\ket{\psi(2)}$ & $\frac{-5}{6\sqrt{3}}$ & $\frac{-7}{18\sqrt{3}}$ & $\frac{-7}{18\sqrt{3}}$ & $\frac{-1}{18\sqrt{3}}$ & $\frac{-1}{18\sqrt{3}}$ & $\frac{-1}{18\sqrt{3}}$ & $\frac{-7}{18\sqrt{3}}$ & $\frac{-5}{6\sqrt{3}}$ & $\frac{-7}{18\sqrt{3}}$ & $\frac{-7}{18\sqrt{3}}$ & $\frac{-5}{6\sqrt{3}}$ & $\frac{-7}{6\sqrt{3}}$ \\[0.7em]
	Vertex Probability & & $\frac{323}{972}$ & & & $\frac{1}{324}$ & & & $\frac{323}{972}$ & & & $\frac{323}{972}$ & \\[0.25em]
	\bottomrule
\end{tabular}
}
\end{center}
\end{table}

\begin{figure}
\begin{center}
	\subfloat[] {
		\includegraphics{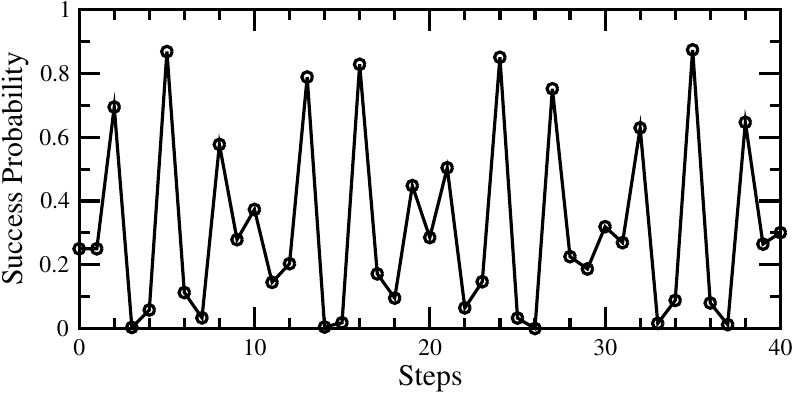}
		\label{fig:quantum_discrete_N4} 
	}

	\subfloat[] {
		\includegraphics{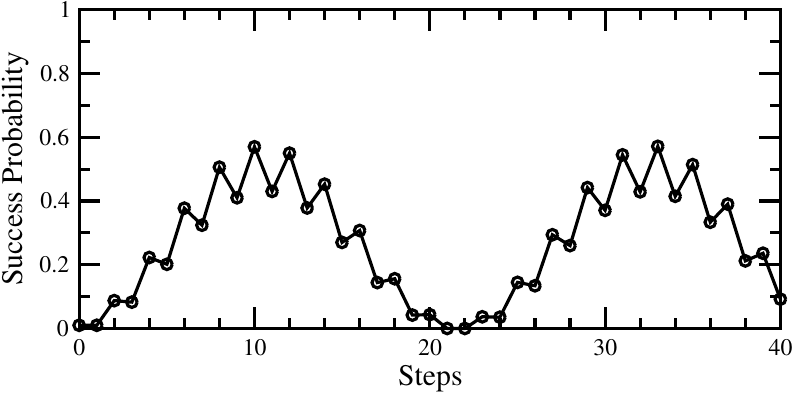}
		\label{fig:quantum_discrete_N100} 
	}
	\vspace*{13pt}
	\fcaption{Success probability for search on the complete graph of (a) $N = 4$ and (b) $N = 100$ vertices by discrete-time quantum walk.}
\end{center}
\end{figure}

Now, let us present the quantum search algorithm by using \fref{fig:K4_directions} as an example, and say vertex $a = 2$ is the marked vertex that we are searching for. Later, we will define the algorithm precisely, but the mathematics is a little complicated, so we will build intuition here. As with random walks, the quantum walk begins in a uniform state $\ket{\psi(0)}$. Since there are twelve amplitudes $\alpha_{ij}$, each amplitude is initially $1/\sqrt{12} = 1/2\sqrt{3}$, as shown in the first row of \tref{table:quantum_discrete}. Using the Born rule, the probability that the walker is found at any vertex is $|1/2\sqrt{3}|^2 + |1/2\sqrt{3}|^2 + |1/2\sqrt{3}|^2 = 3/12 = 1/4$, as expected since there are four vertices. This is listed in the second row of \tref{table:quantum_discrete}. The initial success probability of $1/4$ is plotted as the point $(0,0.25)$ in \fref{fig:quantum_discrete_N4}, where at $0$ steps, the success probability is $0.25$.

To evolve this, one step of the quantum search algorithm consists of querying the oracle, applying a ``coin flip,'' and shifting to adjacent vertices. First, the oracle flips the signs of the amplitudes at the marked vertex. In our example, the marked vertex is $a = 2$, so the amplitudes $\alpha_{21}$, $\alpha_{23}$, and $\alpha_{24}$ get negated to $-1/2\sqrt{3}$, as shown in the third row of \tref{table:quantum_discrete}. Then, we apply a quantum coin flip that inverts each amplitude about the average amplitude at its vertex. That is, at vertex $1$, the average amplitude is $(\alpha_{12} + \alpha_{13} + \alpha_{14}) / 3 = 1/2\sqrt{3}$, and we invert $\alpha_{12}$, $\alpha_{13}$, and $\alpha_{14}$ about this average. In this case, nothing happens. Similarly, at vertex $2$, the average amplitude is $-1/2\sqrt{3}$, and inverting $\alpha_{21}$, $\alpha_{23}$, and $\alpha_{24}$ about this average does nothing. Again, nothing happens for the amplitudes at vertices $3$ and $4$. The result of the coin operation is shown in the fourth row of \tref{table:quantum_discrete}. Finally, we apply a shift operator that causes a walker at vertex $i$ pointing to vertex $j$ to hop to vertex $j$ and point back to vertex $i$. This swaps amplitudes $\alpha_{ij}$ and $\alpha_{ji}$, and the result is shown in the fifth row of \tref{table:quantum_discrete}. This is $\ket{\psi(1)}$, the state of the system after one step of the search algorithm. If we were to measure the position of the walker at this point, we would find it at each vertex with probability $1/4$, as listed in the sixth row of \tref{table:quantum_discrete}. Thus, the success probability after one timestep remains $1/4 = 0.25$, and this point $(1,0.25)$ is plotted in \fref{fig:quantum_discrete_N4}.

\begin{figure}
\begin{center}
	\subfloat[] {
		\includegraphics{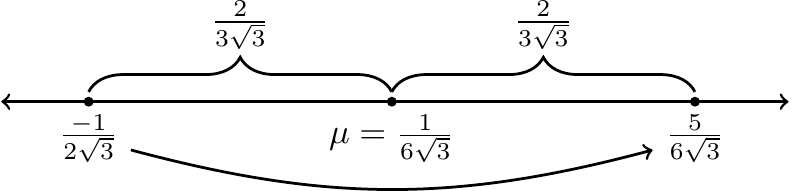}
		\label{fig:coin_reflection_12}
	}

	\subfloat[] {
		\includegraphics{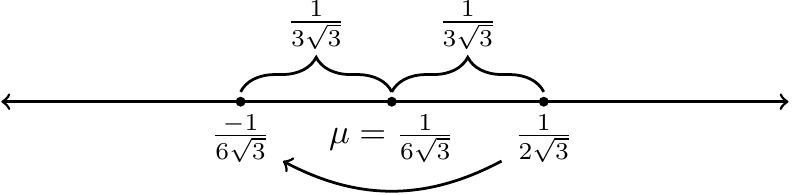}
		\label{fig:coin_reflection_13} 
	}
	\vspace*{13pt}
	\fcaption{For search on the complete graph of $N = 4$ vertices with $a = 2$ marked using a discrete-time quantum walk, the inversions at unmarked vertices for the second application of the Grover diffusion coin.}
\end{center}
\end{figure}

The second step of the search algorithm is much more interesting. The oracle query flips the signs of $\alpha_{21}$, $\alpha_{23}$, and $\alpha_{24}$, as shown in the seventh row of \tref{table:quantum_discrete}. Then, the coin inverts or reflects each amplitude about their average. At vertex $1$, the average amplitude is $\mu = (-1/2\sqrt{3} + 1/2\sqrt{3} + 1/2\sqrt{3})/3 = 1/6\sqrt{3}$. To explain how $-1/2\sqrt{3}$ gets inverted about $\mu$, we plot them on the number line in \fref{fig:coin_reflection_12}. Since $-1/2\sqrt{3}$ is a distance of $2/3\sqrt{3}$ to the left of $\mu$, it gets inverted to that distance to the right of $\mu$, or to $5/6\sqrt{3}$. Similarly, to see how $1/2\sqrt{3}$ gets inverted about $\mu$, we plot them on the number line in \fref{fig:coin_reflection_13}. Since $1/2\sqrt{3}$ is a distance of $1/3\sqrt{3}$ to the right of $\mu$, it gets inverted a distance of $1/3\sqrt{3}$ to the left of $\mu$, or to $-1/6\sqrt{3}$. These values are shown in the eighth row of \tref{table:quantum_discrete}. The same amplitudes are obtained for vertices $3$ and $4$. For vertex $2$, since all the amplitudes are the same, inverting about their average does nothing. Finally, the shift swaps $\alpha_{ij}$ with $\alpha_{ji}$, and the result is shown in the ninth row of \tref{table:quantum_discrete}. This is $\ket{\psi(2)}$, the state of the system after two steps of the search algorithm. If we were to measure the position of the walker, we would find it at vertices $1$, $3$, or $4$, each with probability $|-1/2\sqrt{3}|^2 + |-1/6\sqrt{3}|^2 + |-1/6\sqrt{3}|^2 = 11/108$, or vertex $2$ with probability $|5/6\sqrt{3}|^2 + |5/6\sqrt{3}|^2 + |5/6\sqrt{3}|^2 = 25/36 = 0.69$. This is the tenth row of \tref{table:quantum_discrete}. In \fref{fig:quantum_discrete_N4}, this success probability is plotted as the point $(2,0.69)$, so the probability of the walker being measured at the marked vertex has gone up from its initial value of $0.25$.

The third step of the search algorithm is shown in rows eleven through fourteen of \tref{table:quantum_discrete}, yielding $\ket{\psi(3)}$. Now, the success probability has dropped to $1/324 = 0.003$, and this is plotted as $(3,0.003)$ in \fref{fig:quantum_discrete_N4}. In fact, \fref{fig:quantum_discrete_N4} shows the first forty steps of the algorithm, and the success probability oscillates up and down. With more vertices, however, the oscillations from step-to-step are suppressed, as shown in \fref{fig:quantum_discrete_N100} with $N = 100$ vertices. The idea is to measure the walker when the probability of it being at the marked vertex is high.

Let us now mathematically formalize the discrete-time quantum walk search algorithm. The algorithm was first developed by \cite{SKW2003} for searching the hypercube, and it was generalized to Markov chains by \cite{Szegedy2004}. It was applied to searching the complete graph in \cite{Wong10}, but that focused on the asymptotic behavior of the algorithm. Here, we will also derive the exact evolution. First, let us define the vector space. Each vertex is represented by a basis vector
\[ \ket{1} = \begin{pmatrix} 1 \\ 0 \\ 0 \\ \vdots \\ 0 \end{pmatrix}, \quad \ket{2} = \begin{pmatrix} 0 \\ 1 \\ 0 \\ \vdots \\ 0 \end{pmatrix}, \quad \dots, \quad \ket{N} = \begin{pmatrix} 0 \\ 0 \\ 0 \\ \vdots \\ 1 \end{pmatrix}. \]
Together, these span the vector space $\mathbb{C}^N$, since quantum states are generally superpositions, or complex linear combinations of basis states. Besides this vertex space, there is also an $(N-1)$-dimensional coin space for the $N-1$ directions that the walker can point. Let us denote its basis vectors by
\[ \ket{d_1} = \begin{pmatrix} 1 \\ 0 \\ 0 \\ \vdots \\ 0 \end{pmatrix}, \quad \ket{d_2} = \begin{pmatrix} 0 \\ 1 \\ 0 \\ \vdots \\ 0 \end{pmatrix}, \quad \dots, \quad \ket{d_{N-1}} = \begin{pmatrix} 0 \\ 0 \\ 0 \\ \vdots \\ 1 \end{pmatrix}. \]
Note each of these basis vectors have length $N-1$, in contrast to the vertex-space basis vectors, which had length $N$. Again, since quantum states are generally complex superpositions, the coin space is $\mathbb{C}^{N-1}$. The overall space of the quantum walk is $\mathbb{C}^N \otimes \mathbb{C}^{N-1}$, where $\otimes$ denotes the tensor or Kronecker product. As we saw in the example in \tref{table:quantum_discrete}, this particular algorithm actually maintains real amplitudes throughout its evolution, so the space is actually $\mathbb{R}^N \otimes \mathbb{R}^{N-1}$. This is listed in the second row of \tref{table:summary} (the first row, the oracle, will be discussed later). Then, if the walker is at vertex $i$ and is pointing in the $j$th direction, the basis state is $\ket{i} \otimes \ket{d_j}$. In this tutorial, we encode the directions so that the first direction $\ket{d_1}$ points to lowest numbered vertex, the second direction $\ket{d_2}$ points to the second-lowest numbered vertex, and so forth. More precisely, $\ket{i} \otimes \ket{d_j}$ points to $\ket{j}$ if $j < i$ and to $\ket{j+1}$ if $j \ge i$. To make the direction even more explicit, let us denote a walker at vertex $i$ pointing to vertex $j$ as $\ket{i \to j}$. So, $\ket{i \to j} = \ket{i} \otimes \ket{j}$ if $j < i$ and $\ket{i} \otimes \ket{j-1}$ if $j \ge i$. Using this notation, the general state $\ket{\psi}$ of a quantum walk on the complete graph is a complex linear combination over the possible positions and directions:
\[ \ket{\psi} = \sum_{i = 1}^N \sum_{\substack{j=1 \\ j \ne i}}^N \alpha_{ij} \ket{i \to j} = \begin{pmatrix} \alpha_{12} \\ \alpha_{13} \\ \vdots \\ \alpha_{1N} \\ \alpha_{21} \\ \alpha_{23} \\ \vdots \\ \alpha_{2N} \\ \vdots \\ \alpha_{N1} \\ \alpha_{N2} \\ \vdots \\ \alpha_{N,N-1} \end{pmatrix}. \]
This state vector has length $N(N-1)$, and it reproduces \eqref{eq:quantum_discrete_psi_N4} when $N = 4$.

As we previously described, for measurements, the probability that the walker is found at vertex $i$ pointing to vertex $j$ is equal to $|\alpha_{ij}|^2$. Additionally, if the walker is indeed found at vertex $i$ pointing toward vertex $j$, then the state of the walker after the measurement is simply $\ket{i \to j}$, and we stay the state has ``collapsed'' to this state.

With our vector space defined, initial state is simply a uniform superposition over all the basis states $\ket{i \to j}$:
\begin{equation}
	\label{eq:quantum_discrete_psi0}
	\ket{\psi(0)} = \frac{1}{\sqrt{N(N-1)}} \sum_{i = 1}^N \sum_{\substack{j = 1 \\ j \ne i}}^N \ket{i \to j}.
\end{equation}
This is listed in the third row of \tref{table:summary}. As a check, the norm-square of each amplitude is $1/N(N-1)$, and since there are $N(N-1)$ of them, the total probability is 1, as expected. Alternatively, the probability that the walker will be found at each vertex is $1/N$, as expected.

Whereas the discrete-time random walk evolved by multiplying the state by a stochastic matrix, the discrete-time quantum walk evolves by multiplying the state by a unitary matrix. A matrix is unitary if, when multiplied by its conjugate transpose, the result is the identity matrix. For both walks, stochastic and unitary matrices are respectively required to ensure that the total probability remains 1. Now, since the inverse of a unitary matrix is simply its conjugate transpose, the inverse always exists, so the evolution is reversible. Then, for a quantum walk searching for a marked vertex, amplitude cannot jump to the marked vertex and stay there like a traditional absorbing vertex because this is irreversible. Instead, when a random walk with an absorbing vertex is converted to a quantum walk \cite{Szegedy2004}, we get a phase flip at the marked vertex \cite{Magniez2012,Wong26}. To define it, we first define the uniform superposition over the $N-1$ directions that a particle can point:
\[ \ket{s_c} = \frac{1}{\sqrt{N-1}} \sum_{j=1}^{N-1} \ket{d_j} = \frac{1}{\sqrt{N-1}} \begin{pmatrix} 1 \\ \vdots \\ 1 \end{pmatrix}. \]
The subscript $c$ is to indicate that $\ket{s_c}$ is a state in the coin space. Also denoting $\ket{a,s_c} = \ket{a} \otimes \ket{s_c}$ as a walker at the marked vertex pointing uniformly in all directions, the oracle query is \cite{AKR2005}
\begin{equation}
	\label{eq:quantum_discrete_oracle}
	Q = I_{N(N-1)} - 2 \ketbra{a,s_c}{a,s_c}.
\end{equation}
Here, $I_{N(N-1)}$ is the $N(N-1)$-dimensional identity matrix. Note $Q \ket{a,s_c} = -\ket{a,s_c}$, so the oracle query applies a sign flip to this state. For our particular problem of search the complete graph with a unique marked vertex, the amplitudes at the marked vertex evolve identically. For example, in \tref{table:quantum_discrete} where $N = 4$ and $a = 2$, $\alpha_{21} = \alpha_{23} = \alpha_{24}$ throughout the evolution. Then, the oracle simplify flips the sign of all the amplitudes at the marked vertex:
\begin{align*}
	&\alpha_{21} \ket{2 \to 1} + \alpha_{23} \ket{2 \to 3} + \alpha_{24} \ket{2 \to 4} \\
	&\quad = \alpha_{21} \ket{2} \otimes \ket{d_1} + \alpha_{21} \ket{2} \otimes \ket{d_2} + \alpha_{21} \ket{2} \otimes \ket{d_3} \\
	&\quad = \alpha_{21} \ket{2} \otimes \left( \ket{d_1} + \ket{d_2} + \ket{d_3} \right) \\
	&\quad = \alpha_{21} \sqrt{N-1} \ket{2} \otimes \ket{s_c} \\
	&\quad \xrightarrow{Q} -\alpha_{21} \sqrt{N-1} \ket{2} \otimes \ket{s_c} \\
	&\quad = -\alpha_{21} \ket{2 \to 1} - \alpha_{23} \ket{2 \to 3} - \alpha_{24} \ket{2 \to 4}.
\end{align*}
This proves that $\alpha_{21} \rightarrow -\alpha_{21}$, etc., so the oracle flips the amplitudes at the marked vertex. For any unmarked state $\ket{i \to j}$ with $i \ne a$, using \eqref{eq:quantum_discrete_oracle},
\[ Q \ket{i \to j} = I_{N(N-1)} \ket{i \to j} - 2 \ket{a,s_c} \underbrace{\braket{a,s_c}{i \to j}}_0 = \ket{i \to j}. \]
So, unmarked amplitudes are unchanged by the oracle. Given this behavior of the oracle, another way to write it is
\[ Q' = (I_N - 2 \ketbra{a}{a}) \otimes I_{N-1}, \]
since it inverts the amplitudes of the marked vertex only. This is called a ``phase oracle,'' and it is listed in the first row of \tref{table:summary}. Note $Q'$ is only equal to $Q$ \eqref{eq:quantum_discrete_oracle} when the amplitudes at the marked vertex evolve identically (here, $\alpha_{21} = \alpha_{23} = \alpha_{24}$). Otherwise, $Q$ and $Q'$ are different oracles, and they can lead to different evolutions \cite{Wong10}.

\begin{figure}
\begin{center}
	\includegraphics{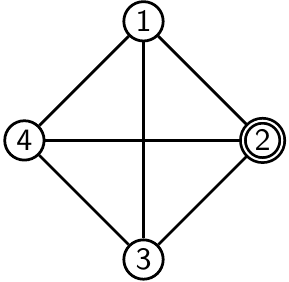}
	\vspace*{13pt}
	\fcaption{\label{fig:K4_marked} The complete graph of $N = 4$ vertices, where vertex $a = 2$ is marked by a quantum oracle, denoted by a double circle.}
\end{center}
\end{figure}

Next, the coin operator is
\[ C = I_N \otimes C_G, \]
where $C_G$ acts on the coin space and is defined by
\[ C_G = 2 \ketbra{s_c}{s_c} - I_{N-1} = \frac{2}{N-1} \begin{pmatrix}
	1 - \frac{N-1}{2} & 1 & 1 & \dots & 1 \\
	1 & 1 - \frac{N-1}{2} & 1 & \dots & 1 \\
	1 & 1 & 1 - \frac{N-1}{2} & \dots & 1 \\
	\vdots & \vdots & \vdots & \ddots & \vdots \\
	1 & 1 & 1 & \dots & 1 - \frac{N-1}{2} \\
\end{pmatrix}, \]
This ``Grover diffusion'' coin \cite{SKW2003} inverts each amplitude about the average amplitude at its vertex (for a proof, see Lemma 2 of \cite{Wong23}). It resembles the ``inversion about the mean'' in Grover's algorithm \cite{Grover1996}, except it only acts on the coin space.

Finally, the ``flip-flop'' shift $S$ causes the walker to hop and then turn around \cite{AKR2005}. So, $S \ket{i \to j} = \ket{j \to i}$.

Note the coin and shift together, \ie, $SC$, is one step of a quantum walk. The search algorithm queries the oracle $Q$ and then takes a step of a quantum walk $SC$, so one step of the search algorithm is
\begin{equation}
	\label{eq:quantum_discrete_evolution}
	\ket{\psi(t+1)} = SCQ \ket{\psi(t)}.
\end{equation}
This evolution is listed in the fourth row of \tref{table:summary}. From the initial state \eqref{eq:quantum_discrete_psi0}, the state at time $t$ is
\begin{equation}
	\label{eq:quantum_discrete_solution}
	\ket{\psi(t)} = (SCQ)^t \ket{\psi(0)}.
\end{equation}
This is the fifth row of \tref{table:summary}.

Graphically, instead of the walk taking place on \fref{fig:K4_absorbing}, where the marked vertex was absorbing, we can imagine the search algorithm taking place on \fref{fig:K4_marked}, where marked vertex is denoted by a double circle \cite{Wong9}. The oracle flips the amplitudes at this marked vertex, and then the quantum walk occurs on the underlying complete graph, where there are no self-loops.

Note the initial state has real amplitudes, and the oracle query, coin, and shift are all real operators. This proves our earlier statement that the search algorithm actually evolves in the space $\mathbb{R}^N \otimes \mathbb{R}^{N-1}$, as listed in the second row of \tref{table:summary}.

Now that all the operators have been precisely defined, we can prove the behavior of the algorithm. Our analysis follow \cite{Wong10}, but that paper focused on the large $N$ behavior of the algorithm, and here we will also derive the exact evolution for all $N$. First, due to the symmetry of the problem, there are only three unique amplitudes: the amplitude of the walker at the marked vertex pointing to an unmarked vertex, the amplitude of the walker at an unmarked vertex pointing to the marked vertex, and the amplitude of the walker at an unmarked vertex pointing at another unmarked vertex. Then, the system evolves in a 3D subspace, as listed in the sixth row of \tref{table:summary}, spanned by:
\begin{align*}
	| ab \rangle &= \frac{1}{\sqrt{N-1}} \sum_{i \ne a} | a \to i \rangle \\
	| ba \rangle &= \frac{1}{\sqrt{N-1}} \sum_{i \ne a} | i \to a \rangle \\
	| bb \rangle &= \frac{1}{\sqrt{N-1}} \sum_{i \ne a} \frac{1}{\sqrt{N-2}} \sum_{\substack{j \ne a \\ j \ne i}} | i \to j \rangle.
\end{align*}
In this $\{ | ab \rangle, | ba \rangle, | bb \rangle \}$ basis, the initial state \eqref{eq:quantum_discrete_psi0} is
\[ | \psi(0) \rangle = \frac{1}{\sqrt{N}} \left( | ab \rangle + | ba \rangle + \sqrt{N-2} | bb \rangle \right). \]
The search operator $U = SCQ$ can be determined by calculating how it acts on each basis vector. This was first worked out in Section 2 of \cite{Wong10}, but more details can be found in Section III.B of \cite{Wong18}. First, with $\ket{ab}$,
\[ U \ket{ab} = SCQ \ket{ab} = -SC \ket{ab} = -S \ket{ab} = -\ket{ba}. \]
Then, with $\ket{ba}$
\begin{align*}
	U \ket{ba} 
		&= SCQ \ket{ba} = SC \ket{ba} = S \left( -\frac{N-3}{N-1} \ket{ba} + \frac{2\sqrt{N-2}}{N-1} \ket{bb} \right) \\
		&= -\frac{N-3}{N-1} \ket{ab} + \frac{2\sqrt{N-2}}{N-1} \ket{bb}.
\end{align*}
Finally, with $\ket{bb}$,
\begin{align*}
	U \ket{bb} 
		&= SCQ \ket{bb} = SC \ket{bb} = S \left( \frac{2\sqrt{N-2}}{N-1} \ket{ba} + \frac{N-3}{N-1} \ket{bb} \right) \\
		&= \frac{2\sqrt{N-2}}{N-1} \ket{ab} + \frac{N-3}{N-1} \ket{bb}.
\end{align*}
Putting these together, the search operator in the $\{ \ket{ab}, \ket{ba}, \ket{bb} \}$ subspace is
\[ U =  \begin{pmatrix}
	0 & -\frac{N-3}{N-1} & \frac{2\sqrt{N-2}}{N-1} \\
	-1 & 0 & 0 \\
	0 & \frac{2\sqrt{N-2}}{N-1} & \frac{N-3}{N-1} \\
\end{pmatrix}. \]
Its (unnormalized) eigenvectors and eigenvalues are
\begin{align*}
	\ket{v_1} &= \frac{1-i\sqrt{2N-3}}{2 \sqrt{N-2}} \ket{ab} + \frac{1+i\sqrt{2N-3}}{2 \sqrt{N-2}} \ket{ba} + \ket{bb}, && \lambda_1 = e^{i\phi}, \\
	\ket{v_2} &= \frac{1+i\sqrt{2N-3}}{2 \sqrt{N-2}} \ket{ab} + \frac{1-i\sqrt{2N-3}}{2 \sqrt{N-2}} \ket{ba} + \ket{bb}, && \lambda_2 = e^{-i\phi}, \\
	\ket{v_3} &= -\sqrt{N-2} \ket{ab} - \sqrt{N-2} \ket{ba} + \ket{bb}, && \lambda_3 = -1,
\end{align*}
where
\[ \sin\phi = \frac{\sqrt{2N-3}}{N-1}, \quad \text{and} \quad \cos\phi = \frac{N-2}{N-1}. \]
Since these eigenvectors span the space, we can express the initial state as a superposition of them:
\[ \ket{\psi(0)} = \frac{\sqrt{N-2}}{(2N-3)\sqrt{N}} \left[ (N-1) \ket{v_1} + (N-1) \ket{v_2} - \ket{v_3} \right]. \]
Then, the state of the system at time $t$ is
\begin{align*}
	\ket{\psi(t)}
		&= U^t \ket{\psi(0)} \\
		&= \frac{\sqrt{N-2}}{(2N-3)\sqrt{N}} \left[ (N-1) U^t \ket{v_1} + (N-1) U^t \ket{v_2} - U^t \ket{v_3} \right] \\
		&= \frac{\sqrt{N-2}}{(2N-3)\sqrt{N}} \left[ (N-1) \lambda_1^t \ket{v_1} + (N-1) \lambda_2^t \ket{v_2} - \lambda_3^t \ket{v_3} \right] \\
		&= \frac{\sqrt{N-2}}{(2N-3)\sqrt{N}} \left[ (N-1) e^{i\phi t} \ket{v_1} + (N-1) e^{-i\phi t} \ket{v_2} - (-1)^t \ket{v_3} \right] \\
		&= \frac{\sqrt{N-2}}{(2N-3)\sqrt{N}} \\
		&\quad \times \bigg\{ \bigg[ \frac{N-1}{\sqrt{N-2}} \left( \cos(\phi t) + \sin(\phi t) \sqrt{2N-3}\right) + (-1)^t \sqrt{N-2} \bigg] \ket{ab} \\
		&\quad\quad + \left[ \frac{N-1}{\sqrt{N-2}} \left( \cos(\phi t) - \sin(\phi t) \sqrt{2N-3} \right) + (-1)^t \sqrt{N-2} \right] \ket{ba} \\
		&\quad\quad + \left[ 2(N-1) \cos(\phi t) + (-1)^{t+1} \right] \ket{bb} \bigg\},
\end{align*}
where in the last equality, we plugged in for $\ket{v_1}$, $\ket{v_2}$, and $\ket{v_3}$ and then used $e^{i\phi t} + e^{-i\phi t} = 2 \cos(\phi t)$ and $e^{i\phi t} - e^{-i\phi t} = 2i \sin(\phi t)$. Then, the success probability at time $t$ is the norm-square of the coefficient or amplitude of $\ket{ab}$:
\begin{equation}
	\label{eq:quantum_discrete_prob_time}
	\frac{1}{(2N-3)^2N} \left[ (N-1) \left( \cos(\phi t) + \sin(\phi t) \sqrt{2N-3} \right) + (-1)^t (N-2) \right]^2.
\end{equation}
This is a new result, and it is listed in the seventh row of \tref{table:summary}. As a check, it exactly replicates \fref{fig:quantum_discrete_N4} when $N = 4$ and \fref{fig:quantum_discrete_N100} when $N = 100$.

To maximize the success probability, we want the number of steps to be
\begin{equation}
	\label{eq:quantum_discrete_runtime}
	t = \frac{\pi}{2\phi}.
\end{equation}
This causes $\cos(\phi t) = 0$ and $\sin(\phi t) = 1$. Of course, $t$ can only take integer values since the evolution is discrete in time, so we must round. Let us round to the nearest even integer so that $(-1)^t = 1$. Then, it may not be true that $\cos(\phi t) = 0$ and $\sin(\phi t) = 1$, but we can bound their error. If we round $t$ to the nearest even integer, then $t$ is within 1 from $\pi/2\phi$, i.e.,
\[ \frac{\pi}{2\phi} - 1 \le t \le \frac{\pi}{2\phi} + 1. \]
Multiplying by $\phi$,
\[ \frac{\pi}{2} - \phi \le \phi t \le \frac{\pi}{2} + \phi. \]
Then, $\cos(\phi t)$ is bounded by
\[ \cos\left( \frac{\pi}{2} \pm \phi \right) = \mp \phi \pm \frac{\phi^3}{6} \mp \dots = O(\phi) = O \left( \frac{1}{\sqrt{N}} \right), \]
where in the last inequality, we used for large $N$,
\[ \phi = \sin^{-1} \left( \frac{\sqrt{2N-3}}{N-1} \right) \approx \sqrt{\frac{2}{N}}. \]
Similarly, $\sin(\phi t)$ is bounded by
\[ \sin\left( \frac{\pi}{2} \pm \phi \right) = 1 - \frac{\phi^2}{2} + \dots = 1 + O \left( \phi^2 \right) = 1 + O \left( \frac{1}{N} \right). \]
Using these and $(-1)^t = 1$, the success probability \eqref{eq:quantum_discrete_prob_time} becomes
\[ \frac{1}{(2N-3)^2N} \left[ (N-1) \left[ O \left( \frac{1}{\sqrt{N}} \right) + \left( 1 + O \left( \frac{1}{N} \right) \right) \sqrt{2N-3} \right] + N-2 \right]^2. \]
Distributing and simplifying, this becomes
\[ \frac{1}{(2N-3)^2N} \left[ (N-1) \sqrt{2N-3} + N-2 + O \left( \sqrt{N} \right) \right]^2. \]
In the square bracket, we Taylor expand $(N-1)\sqrt{2N-3} + N - 2$ for large $N$ and only keep terms $O(\sqrt{N})$, resulting in
\[ \frac{1}{(2N-3)^2N} \left[ \sqrt{2} N^{3/2} + N + O \left( \sqrt{N} \right) \right]^2. \]
Pulling out the big-$O$, it becomes $O(N^{3/2} \sqrt{N})$, but dividing by the overall factor, it becomes $O(1/N)$:
\[ \frac{1}{(2N-3)^2N} \left[ \sqrt{2} N^{3/2} + N \right]^2 + O \left( \frac{1}{N} \right). \]
Simplifying, the success probability is
\begin{equation}
	\label{eq:quantum_discrete_prob}
	\frac{N(\sqrt{2N}+1)^2}{(2N-3)^2} + O \left( \frac{1}{N} \right).
\end{equation}
This is a new result, and it is listed in the eighth row of \tref{table:summary}. As a numerical check, when $N = 100$, \eqref{eq:quantum_discrete_prob} yields a success probability of $0.59$, and this closely matches \fref{fig:quantum_discrete_N100}. The runtime \eqref{eq:quantum_discrete_runtime} is also listed in the eight row of \tref{table:summary}, and it should be rounded to the nearest even integer. 

Next, for further asymptotic results, plugging $\phi \approx \sqrt{2/N}$ into \eqref{eq:quantum_discrete_runtime} yields the asymptotic runtime
\begin{equation}
	\label{eq:quantum_discrete_runtime_largeN}
	\frac{\pi}{2\sqrt{2}} \sqrt{N}.
\end{equation}
The success probability at this time is \eqref{eq:quantum_discrete_prob}, which for large $N$ is
\begin{equation}
	\label{eq:quantum_discrete_prob_largeN}
	\frac{1}{2} + O \left( \frac{1}{\sqrt{N}} \right).
\end{equation}
These asymptotic behaviors are listed, without the big-$O$, in the ninth row of \tref{table:summary}, and they were previously known from \cite{Wong10}. Since the success probability asymptotically reaches $1/2$, we expect to repeat the algorithm twice, on average, before finding the marked vertex. Then, the overall expected runtime is $\pi\sqrt{N}/\sqrt{2} = O(\sqrt{N})$, as listed in the final row of \tref{table:summary}. A more rigorous way to understand repeating the algorithm is to use basic probability theory. If we run the algorithm $r$ times, then the probability that we still have not found the marked vertex is $1/2$ multiplied by itself $r$ times, or $1/2^r$. Taking the complement, if we run the algorithm $r$ times, the probability of finding the marked vertex is $1 - 1/2^r$. Changing the notation so that it is similar to our analysis of the (classical) random walks, let us define $\epsilon = 1/2^r$. Then, running the algorithm $r$ times, the probability of finding the marked vertex is
\[ 1 - \epsilon, \]
and the total number of queries is \eqref{eq:quantum_discrete_runtime_largeN} times $r = \log_2(1/\epsilon)$, i.e.,
\[ \frac{\pi}{2\sqrt{2}} \sqrt{N} \log_2 \left( \frac{1}{\epsilon} \right). \]
These are also listed in the ninth row of \tref{table:summary}.


\section{Continuous-Time Quantum Walk}

Finally, we explore how a continuous-time quantum walk searches the complete graph for a marked vertex, corresponding to the rightmost column of \tref{table:summary}. Continuous-time quantum walks were introduced in \cite{FG1998b} to solve decision trees, and they were first applied to spatial search problems in \cite{CG2004}, including search on the complete graph. Ref.~\cite{CG2004} showed that a continuous-time quantum walk searching the complete graph is equivalent the ``analogue analog'' of Grover's algorithm proposed in \cite{FG1998a}, up to a global phase. Thus, the following analysis of how a continuous-time quantum walk searches the complete graph is equivalent to \cite{FG1998a}, up to a global phase. Our results are also similar to \cite{Wong10}, up to a global phase, but that paper only considered what happened at the runtime. In contrast, here we derive the success probability at all time $t$, like \cite{FG1998a} did.

As before, the vertices are the basis vectors $\ket{1}, \ket{2}, \dots, \ket{N}$ of an $N$-dimensional space. Unlike a discrete-time quantum walk, a continuous-time quantum walk does not need the coin degree of freedom to evolve nontrivially. So, the space of the quantum walk is $\mathbb{C}^N$, and this is listed in the second row of \tref{table:summary} (we will get to the first row, the type of oracle, later). Then, the state of a continuous-time quantum walk is simply a superposition, or complex linear combination, of the vertices, \ie,
\[ \ket{\psi} = \alpha_1 \ket{1} + \alpha_2 \ket{2} + \dots + \alpha_N \ket{N} = \begin{pmatrix} \alpha_1 \\ \alpha_2 \\ \vdots \\ \alpha_N \end{pmatrix}, \]
where $\alpha_i \in \mathbb{C}$. Following the Born rule, when measured, the probability that the walker will be found at vertex $i$ is given by $|\alpha_i|^2$. Then, since the total probability is 1, $\sum_i |\alpha_i|^2 = 1$. If the walker is found at vertex $i$, then the state of the system after the measurement has collapsed to $\ket{i}$.

As with the random walks, the initial state of the quantum walk is uniform over the vertices:
\begin{equation}
	\label{eq:quantum_continuous_psi0}
	\ket{\psi(0)} = \frac{1}{\sqrt{N}} \sum_{i=1}^N \ket{i}.
\end{equation}
This is the third row of \tref{table:summary}. As a check, if we were to measure the position of the walker at the start, the probability of finding it at each vertex is $|1/\sqrt{N}|^2 = 1/N$, which is a uniform probability distribution.

This state evolves by Schr\"odinger's equation, which is the fundamental equation of quantum mechanics \cite{Griffiths2018}. In units where $\hbar$, which is Planck's constant divided by $2\pi$, is equal to 1, Schr\"odinger's equation is
\begin{equation}
	\label{eq:quantum_continuous_evolution}
	i \frac{d\ket{\psi}}{dt} = H \ket{\psi},
\end{equation}
where $i = \sqrt{-1}$, and $H$ is the Hamiltonian of the system, which is an operator corresponding to the total energy of the system. As such, the Hamiltonian contains a kinetic energy term $T$ and a potential energy term $V$:
\[ H = T + V. \]
Additionally, the Hamiltonian must be a Hermitian or self-adjoint operator, meaning it is equal to its conjugate transpose. This is so that the total probability of the state stays $1$. Now, if the Hamiltonian is constant in time (\ie, it is time-independent), the solution to Schr\"odinger's equation is
\begin{equation}
	\label{eq:quantum_continuous_solution}
	\ket{\psi(t)} = e^{-iHt} \ket{\psi(0)}.
\end{equation}
Note the similarity of Schr\"odinger's equation and its solution to the continuous-time random walk in \eqref{eq:classical_continuous_evolution} and \eqref{eq:classical_continuous_solution}. We see that to go from the random walk to the quantum walk, we replace the governing matrix from one whose columns summed to zero to a matrix that is Hermitian, and we include the imaginary unit $i$.

From quantum mechanics \cite{Griffiths2018}, the kinetic energy of the walker is 
\[ T = \frac{-1}{2m} \nabla^2, \]
where $m$ is the mass of the walker. As with the continuous-time random walk, we replace the continuous-space Laplace operator $\nabla^2$ with the discrete-space Laplacian $L = A - D$. Also letting $\gamma = 1/2m$, the kinetic energy is
\[ T = -\gamma L. \]
Note $L$ is the Laplacian of the underlying complete graph, so it is Hermitian. It is not \eqref{eq:classical_L}, which had an absorbing vertex, since an absorbing vertex would make $L$ not Hermitian. For the underlying complete graph of $N$ vertices, the Laplacian has elements
\[ L_{ij} = \begin{cases}
	-(N-1), & \text{if } i = j, \\
	1, & \text{if } i \ne j. \\
\end{cases} \]
Since kinetic energy is the energy of motion, it is what effects the quantum walk. The more kinetic energy the walker has, the more quickly it moves. Hence, $\gamma$ can be interpreted as the amplitude per time of the walk.

Rather than acting as an absorbing vertex, the oracle appears in the potential energy term $V$ as
\[ V = -\ketbra{a}{a}. \]
This is a Hamiltonian oracle \cite{Mochon2007}, and it acts like a phase oracle because if $H = V$, then according to \eqref{eq:quantum_continuous_solution}, a vertex $\ket{x}$ evolves by
\[ e^{-iHt} \ket{x} = e^{i \ketbra{a}{a} t} \ket{x} = \begin{cases}
	e^{it} \ket{x}, & x = a, \\
	\ket{x}, & x \ne a.
\end{cases} \]
Thus, the marked vertex evolves by a phase, and unmarked vertices do not evolve at all. As such, the first row of \tref{table:summary} lists the oracle as a Hamiltonian phase oracle.

Adding the kinetic and potential energy terms, the Hamiltonian of the quantum search algorithm is
\begin{equation}
	\label{eq:quantum_continuous_Hamiltonian}
	H = -\gamma L - \ketbra{a}{a}.
\end{equation}
Again, the kinetic energy term effects the walk, and the potential energy encodes the oracle. Notice $\norm{V} = 1$, so however much time the system evolves by this Hamiltonian is the amount of time that the oracle is queried. Then, the goal is to find $\ket{a}$ by evolving with a little time as possible. If we plug the Hamiltonian \eqref{eq:quantum_continuous_Hamiltonian} into \eqref{eq:quantum_continuous_evolution}, we get the fourth row of \tref{table:summary}, and if we plug it into \eqref{eq:quantum_continuous_solution}, we get the fifth row of \tref{table:summary}.

\begin{figure}
\begin{center}
	\includegraphics{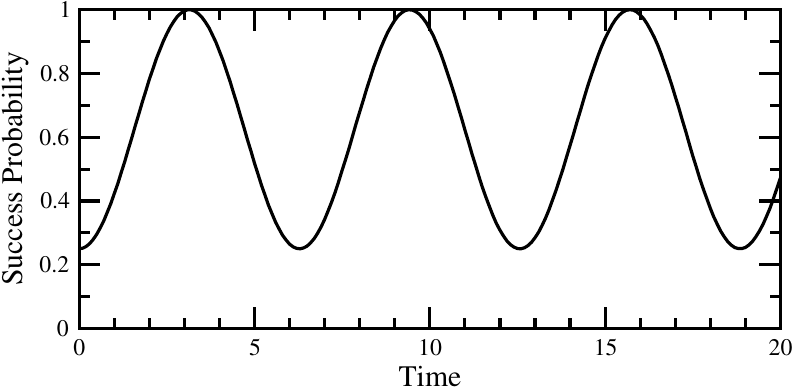}
	\vspace*{13pt}
	\fcaption{\label{fig:quantum_continuous_N4} Success probability for search on the complete graph of $N = 4$ by continuous-time quantum walk.}
\end{center}
\end{figure}

Before we prove the behavior of the algorithm, let us consider an example. For the complete graph of $N = 4$ vertices with $a = 2$ marked, as shown in \fref{fig:K4_marked}, we have
\[ 
	L = \begin{pmatrix}
		-3 & 1 & 1 & 1 \\
		1 & -3 & 1 & 1 \\
		1 & 1 & -3 & 1 \\
		1 & 1 & 1 & -3 \\
	\end{pmatrix}, \quad \ketbra{a}{a} = \begin{pmatrix}
		0 & 0 & 0 & 0 \\
		0 & 1 & 0 & 0 \\
		0 & 0 & 0 & 0 \\
		0 & 0 & 0 & 0 \\
	\end{pmatrix}.
\]
If we choose $\gamma = 1/4$ (more on this later), the Hamiltonian \eqref{eq:quantum_continuous_Hamiltonian} is
\[ H = \frac{1}{4} \begin{pmatrix}
	 3 & -1 & -1 & -1 \\
	-1 & -1 & -1 & -1 \\
	-1 & -1 &  3 & -1 \\
	-1 & -1 & -1 &  3 \\
\end{pmatrix}. \]
Then, using \eqref{eq:quantum_continuous_solution}, we can calculate the state of the system at time $t$. Squaring the amplitude at the marked vertex yields the success probability, which we plot in \fref{fig:quantum_continuous_N4} as a function if time. It starts at $1/4 = 0.25$ and oscillates to 1 and back, periodically. The idea is to measure the walker when the probability of it being at the marked vertex is high.

To analytically prove the behavior of this algorithm, we follow \cite{Wong10}, although as previously noted, its roots are in \cite{CG2004,FG1998a}. Due to the symmetry of the initial state and evolution, the amplitudes at all the unmarked vertices are equal. So, the system evolves in a 2D subspace spanned by $\ket{a}$ and 
\[ \ket{b} = \frac{1}{\sqrt{N-1}} \sum_{i \ne a} \ket{i}. \]
That the system evolves the 2D subspace $\mathbb{C}^2$ is listed the sixth row of \tref{table:summary}. In this $\{ \ket{a}, \ket{b} \}$ basis, the initial state is
\[ \ket{\psi(0)} = \frac{1}{\sqrt{N}} \ket{a} + \sqrt{\frac{N-1}{N}} \ket{b}. \]
Next, to find the Hamiltonian in this basis, we have
\begin{align*}
	H \ket{a} 
		&= \left( -\gamma L - \ketbra{a}{a} \right) \ket{a} \\
		&= -\gamma A \ket{a} + \gamma D \ket{a} - \ket{a} \braket{a}{a} \\
		&= -\gamma \sqrt{N-1} \ket{b} + \gamma (N-1) \ket{a} - \ket{a} \\
		&= \left[ \gamma(N-1) - 1 \right] \ket{a} - \gamma \sqrt{N-1} \ket{b}.
\end{align*}
We also have
\begin{align*}
	H \ket{b} 
		&= \left( -\gamma L - \ketbra{a}{a} \right) \ket{b} \\
		&= -\gamma A \ket{b} + \gamma D \ket{b} - \ket{a} \braket{a}{b} \\
		&= -\gamma \left[ \sqrt{N-1} \ket{a} + (N-2) \ket{b} \right] + \gamma(N-1) \ket{b} - 0 \\
		&= -\gamma \sqrt{N-1} \ket{a} + \gamma \ket{b}.
\end{align*}
So,
\[ H = \begin{pmatrix}
	\gamma(N-1) - 1 & -\gamma \sqrt{N-1} \\
	-\gamma \sqrt{N-1} & \gamma \\
\end{pmatrix}. \]
In \cite{Wong5}, a similar $2 \times 2$ matrix was found for the Hamiltonian $-\gamma N \ketbra{s}{s} - \ketbra{a}{a}$, where $\ket{s} = \sum_i \ket{i}/\sqrt{N}$, and in \cite{Wong10}, a similar matrix was found for the Hamiltonian $-\gamma A - \ketbra{a}{a}$. As shown in \cite{CG2004,Wong5}, $\gamma$ must be chosen to be $1/N$ for the search algorithm to appreciably evolve from its initial state. Then, with $\gamma = 1/N$, the Hamiltonian is
\[ H = \frac{1}{N} \begin{pmatrix}
	-1 & -\sqrt{N-1} \\
	-\sqrt{N-1} & 1 \\
\end{pmatrix}. \]
The (unnormalized) eigenvectors and eigenvalues of this are
\begin{alignat*}{3}
	\ket{v_+} &= \frac{-\sqrt{N-1}}{\sqrt{N} + 1} \ket{a} + \ket{b}, \quad &&E_+ = \frac{1}{\sqrt{N}}, \\
	\ket{v_-} &= \frac{\sqrt{N-1}}{\sqrt{N} - 1} \ket{a} + \ket{b}, \quad &&E_- = \frac{-1}{\sqrt{N}}.
\end{alignat*}
Writing the initial state as a linear combination of these eigenvectors,
\[ \ket{\psi(0)} = \frac{1}{2} \sqrt{\frac{N-1}{N}} \ket{v_+} + \frac{1}{2} \sqrt{\frac{N-1}{N}} \ket{v_-}. \]
Then, using \eqref{eq:quantum_continuous_evolution}, the state of the system at time $t$ is
\begin{align*}
	\ket{\psi(t)} 
		&= e^{-iHt} \ket{\psi(0)} \\
		&= \frac{1}{2} \sqrt{\frac{N-1}{N}} e^{-iHt} \ket{v_+} + \frac{1}{2} \sqrt{\frac{N-1}{N}} e^{-iHt} \ket{v_-} \\
		&= \frac{1}{2} \sqrt{\frac{N-1}{N}} e^{-iE_+t} \ket{v_+} + \frac{1}{2} \sqrt{\frac{N-1}{N}} e^{-iE_-t} \ket{v_-} \\
		&= \frac{1}{2} \sqrt{\frac{N-1}{N}} e^{-it/\sqrt{N}} \left( \frac{-\sqrt{N-1}}{\sqrt{N} + 1} \ket{a} + \ket{b} \right) \\
		&\quad+ \frac{1}{2} \sqrt{\frac{N-1}{N}} e^{it/\sqrt{N}} \left( \frac{\sqrt{N-1}}{\sqrt{N} - 1} \ket{a} + \ket{b} \right) \\
		&= \frac{1}{2} \left[ -\frac{\sqrt{N}-1}{\sqrt{N}} e^{-it/\sqrt{N}} + \frac{\sqrt{N}+1}{\sqrt{N}} e^{it/\sqrt{N}} \right] \ket{a} \\
		&\quad+ \frac{1}{2} \sqrt{\frac{N-1}{N}} \left( e^{it/\sqrt{N}} + e^{-it/\sqrt{N}} \right) \ket{b} \\
		&= \left[ i \sin \left( \frac{t}{\sqrt{N}} \right) + \frac{1}{\sqrt{N}} \cos \left( \frac{t}{\sqrt{N}} \right) \right] \ket{a} \\
		&\quad + \sqrt{\frac{N-1}{N}} \cos \left( \frac{t}{\sqrt{N}} \right) \ket{b}.
\end{align*}
Up to a global phase, this is equivalent to Eq.~(10) of \cite{FG1998a}. Taking the norm-square of the coefficient of $\ket{a}$, the success probability at time $t$ is
\begin{equation}
	\label{eq:quantum_continuous_prob_time}
	\sin^2 \left( \frac{t}{\sqrt{N}} \right) + \frac{1}{N} \cos^2 \left( \frac{t}{\sqrt{N}} \right).
\end{equation}
This is Eq.~(11) of \cite{FG1998a}, and it exactly reproduces \fref{fig:quantum_continuous_N4}, and it is listed in the seventh row of \tref{table:summary}.

From \eqref{eq:quantum_continuous_prob_time}, we see that the success probability reaches $1$ at time
\[ t = \frac{\pi}{2} \sqrt{N}, \]
and this is the eighth row of \tref{table:summary}. This is known from Eq.~(12) of \cite{FG1998a}, Section III.A of \cite{CG2004}, and Section 3 of \cite{Wong10}. It is also the asymptotic behavior, so it is listed again in the ninth row of \tref{table:summary}. Finally, in big-O notation, the overall runtime is $O(\sqrt{N})$, and this is listed in the final row of \tref{table:summary}.

We previously showed that the discrete- and continuous-time random walks converged to the same evolution for large $N$. Now, let us compare the asymptotic behaviors of the quantum walks. From \eqref{eq:quantum_discrete_prob_time}, for large $N$, the discrete-time quantum walk's success probability at time $t$ is
\[ \frac{1}{2} \sin^2 \left( \frac{\sqrt{2} t}{\sqrt{N}} \right), \]
whereas from \eqref{eq:quantum_continuous_prob_time}, the continuous-time quantum walk's success probability at time $t$ is
\[ \sin^2 \left( \frac{t}{\sqrt{N}} \right). \]
These asymptotic evolutions differ in two places: the overall factor of $1/2$, and the factor of $\sqrt{2}$ for time. The overall factor of $1/2$ can actually be eliminated using the fact that the success probability of the discrete-time quantum walk can be doubled by using the direction of the discrete-time quantum walk \cite{Wong18}. That is, our derivation of the success probability only included the norm-square of the amplitude at $\ket{ab}$, since that is the standard treatment. It was proved in \cite{Wong18}, however, that the probability of the state $\ket{ba}$ is asymptotically the same as the probability of $\ket{ab}$. Since $\ket{ba}$ points to the marked vertex, it infers the location of the marked vertex. This doubles the success probability of the discrete-time quantum walk. Then, the only asymptotic difference between the quantum walks is that the discrete-time algorithm evolves more quickly in time by a factor of $\sqrt{2}$. This comparison between the discrete- and continuous-time quantum walks is new.


\section{Conclusion}

In this tutorial, we explored how discrete- and continuous-time random walks and quantum walks solve the unstructured search problem, which is spatial search on the complete graph. The results were summarized in \tref{table:summary}. For the random walks, while the overall runtimes of $O(N)$ were known, the other details seem to be new. We proved that their exact evolutions differ, but for large $N$, their evolutions converge to each other, and they both reach a success probability of $1-\epsilon$ in time $N \ln(1/\epsilon)$. For the quantum walks, the large $N$ behavior of the discrete-time quantum walk was already known, but the exact evolution was new. For the continuous-time quantum walk, the results were known from the ``analog analogue'' of Grover's algorithm \cite{FG1998a}. Asymptotically, although the discrete-time quantum walk reaches half the success probability as the continuous-time quantum walk, it can be doubled to reach the same probability \cite{Wong18}. Then, the quantum walks evolve in the same asymptotic manner, except the discrete-time quantum walk evolves more quickly by a factor of $\sqrt{2}$. Overall, the quantum walks search in $O(\sqrt{N})$ time, which is a quadratic speedup over the random walks.

Finally, we would be remiss if we did not mention that there exist other kinds of quantum walks, such as scattering quantum walks \cite{Hillery2003}, interpolating quantum walks \cite{Krovi2016} and staggered quantum walks \cite{Portugal2016}, but they are beyond the scope of this tutorial.


\nonumsection{Acknowledgements}
\noindent
Thanks to Zak Webb for pointing out the spectral norm in the continuous-time random walk. The early stages of this work were carried out while the author was a postdoctoral scholar at the University of Texas at Austin under Scott Aaronson, so this work was partially supported by the U.S.~Department of Defense Vannevar Bush Faculty Fellowship of Scott Aaronson.


\nonumsection{References}
\noindent
\bibliography{refs}
\bibliographystyle{qic}

\end{document}